\documentclass[twocolumn,aps,prc,floatfix,superscriptaddress]{revtex4-2}

\usepackage{graphicx}
\usepackage{dcolumn}
\usepackage{bm}
\usepackage{hyperref}
\usepackage{amsmath,bm}
\usepackage{amssymb}
\usepackage{longtable}
\usepackage{placeins}
\usepackage[mathlines]{lineno}
\usepackage{xcolor}
\usepackage{upgreek}
\allowdisplaybreaks

\newcommand{\msplast}{m_{\mathrm{n}\text{-}\mathrm{p}}^{\ast}}
\newcommand{\mspl}{m_{\mathrm{n}\text{-}\mathrm{p}}}

\usepackage[normalem]{ulem}
\usepackage{bbding}
\usepackage{booktabs}

\renewcommand{\sout}{\bgroup \color{red} \ULdepth=-.5ex \ULset}

\begin{document}

\title{Extended momentum-dependent interaction for transport models and neutron stars}
% \thanks{version 1}

\author{Si-Pei Wang}
\affiliation{
State Key Laboratory of Dark Matter Physics, Key Laboratory for Particle Astrophysics and Cosmology (MOE),
and Shanghai Key Laboratory for Particle Physics and Cosmology,
School of Physics and Astronomy, Shanghai Jiao Tong University, Shanghai 200240, China
}
\author{Lie-Wen Chen}
 \email{Corresponding author: lwchen@sjtu.edu.cn}
\affiliation{
State Key Laboratory of Dark Matter Physics, Key Laboratory for Particle Astrophysics and Cosmology (MOE),
and Shanghai Key Laboratory for Particle Physics and Cosmology,
School of Physics and Astronomy, Shanghai Jiao Tong University, Shanghai 200240, China
}

\date{\today}

\begin{abstract}
The momentum-dependent interaction (MDI) model, which has been widely used in microscopic transport models for heavy-ion collisions (HICs), is extended to include three different momentum-dependent (MD) terms and three zero-range density-dependent (DD) terms, dubbed as MDI3Y model.
Compared to the MDI model, the single-nucleon potential in the MDI3Y model exhibits more flexible MD behaviors, e.g.,
(1) the nucleon optical model potential $U_0$ in symmetric nuclear matter can have very different high energy behaviors while nicely describe the known data at lower energies;
(2) the isoscalar nucleon effective mass $m_{s}^{\ast}$ can have a peak structure at subsaturation densities;
(3) the nuclear symmetry potential $U_{\mathrm{sym}}$ can display a non-monotonic momentum dependence.
Furthermore,
the inclusion of three zero-range DD interactions follows the idea of Fermi momentum expansion, allowing
more flexible variation for the largely uncertain high-density behaviors of nuclear matter equation of state (EOS), especially the symmetry energy.
Moreover, we also obtain
the corresponding Skyrme-like energy density functional through density matrix expansion of the finite-range exchange interactions.
Based on the MDI3Y model,
we construct four interactions with the same symmetry energy slope parameter $L=35$~MeV but different momentum dependence of $U_{\mathrm{sym}}$, by fitting the empirical nucleon optical potential, the empirical properties of symmetric nuclear matter, the microscopic calculations of pure neutron matter EOS and the astrophysical constraints on neutron stars.
In addition,
two interactions with $L=55$ and $75$~MeV are also constructed for comparison.
Using these MDI3Y interactions, we study the properties of nuclear matter and neutron stars.
These MDI3Y interactions,
especially those with non-monotonic momentum dependence of $U_{\mathrm{sym}}$, will be potentially useful in transport model analyses of HICs data to extract nuclear matter EOS and the isospin splitting of nucleon effective masses.

\end{abstract}

\maketitle

\section{INTRODUCTION}
The EOS of nuclear matter, usually defined as energy (or pressure) vs density, is of fundamental importance in
nuclear physics and astrophysics~\cite{Danielewicz:2002pu,Lattimer:2004pg,Oertel:2016bki}.
Theoretically, it is still a big challenge to calculate the nuclear
matter EOS from first principle, especially at
suprasaturation densities, due to the complicated nonperturbative feature of
QCD~\cite{Brambilla:2014jmp},
although progress in microscopic many-body calculations, see, e.g., Refs.~\cite{Tews:2012fj,Wellenhofer:2015qba,Drischler:2020hwi,Gandolfi:2011xu,Wlazlowski:2014jna,Roggero:2014lga,Tews:2015ufa,Akmal:1998cf,Baldo:2014rda,Carbone:2014mja}, has improved the predictions of the EOS of pure neutron matter~(PNM) (main limited at lower densities).
Therefore,
our knowledge on nuclear matter EOS so far obtained is mainly from model analyses on data from terrestrial experiments or astrophysical observations.
In the past decades, by analyzing
the giant resonances of finite nuclei~\cite{Blaizot:1980tw,Youngblood:1999zza,Shlomo:2006ole,Li:2007bp,Garg:2018uam,Li:2022suc}, as well as the collective flows~\cite{Danielewicz:2002pu,LeFevre:2015paj,Cozma:2024cwc} and the kaon production~\cite{Aichelin:1985rbt,Fuchs:2000kp,Hartnack:2005tr,Fuchs:2005zg} in HICs, the EOS of isospin symmetric nuclear matter (SNM) has been relatively well constrained even up to about five times the nuclear saturation density $\rho_0$.
On the other hand, for the isospin-dependent part of nuclear matter EOS, referred to as the symmetry energy $E_{\mathrm{sym}}$,
its density behavior below and around $\rho_0$ has been
reasonably well constrained by analyzing finite nuclei properties~\cite{Chen:2005ti,Centelles:2008vu,Chen:2010qx,Agrawal:2012pq,Zhang:2013wna,Brown:2013mga,Roca-Maza:2012uor,Zhang:2014yfa,Danielewicz:2013upa,Zhang:2015ava,Danielewicz:2016bgb,Xu:2020xib,Qiu:2023kfu} and observables in HICs at energies less than about $100$~MeV/nucleon~\citep{Chen:2004si,Li:2005jy,Kowalski:2006ju,Shetty:2005qp,Tsang:2008fd,Wada:2011qm,Morfouace:2019jky,Zhang:2020azr}, with its suprasaturation density behavior
remaining among the most uncertain properties of nuclear matter (see, e.g., Ref.~\cite{Sorensen:2023zkk} for a recent review).

In terrestrial labs,
the intermediate and high energy HICs induced by neutron-rich nuclei provides a unique tool to explore the suprasaturation density behaviors of the $E_{\mathrm{sym}}$, as the dense neutron-rich nuclear matter may be formed during the collisions. In nature, the neutron stars and their mergers offer ideal sites to study the high-density $E_{\mathrm{sym}}$ since the interior of neutron stars are dominated by dense and extremely neutron-rich matter.
To investigate the HICs and neutron star properties, one usually adopts the microscopic transport models and energy density functionals based on the same in-medium nuclear effective interactions, which generally depend on the medium baryon
density and isospin asymmetry as well as the momentum and isospin of the interacting particles.
Therefore, the in-medium nuclear effective interactions play a decisive role in extracting physics information from analyzing data of HICs and multimessenger signals of neutron stars and their mergers. In this work, we mainly focus on the momentum dependence of the in-medium nuclear effective interactions.

Microscopic transport models such as the Boltzmann-Uehling-Uhlenbeck (BUU) equation~\cite{Bertsch:1988ik} and quantum molecular dynamics (QMD) model~\cite{Aichelin:1991xy} have been developed to describe the dynamics of HICs and to extract the EOS of dense nuclear matter from the collision observables.
In recent years, the Transport Model Evaluation Project has been comparing the predictions from different models under controlled conditions, and aims to extract reliable constraints on the EOS from HICs~\cite{TMEP:2016tup,TMEP:2017mex,TMEP:2019yci,TMEP:2021ljz,TMEP:2022xjg,TMEP:2023ifw}.
The basic input in the one-body transport models is the single-nucleon potential (nuclear mean-field potential)~\cite{Bertsch:1988ik}, which generally depends on the baryon density and the isospin asymmetry of the medium as well as the isospin and momentum of the particle and can be obtained from the in-medium nuclear effective interactions~\cite{Behera:1979XXX,Decharge:1979fa,Wiringa:1988jt,Behera:1998XXX}.
From the Brueckner theory, the momentum dependence mainly arises from the non-locality of the exchange interaction~\cite{Brueckner:1955zze,Jeukenne:1976uy,Mahaux:1985zz,Bertsch:1988ik,Li:2018lpy}.
Experimentally, the momentum dependence of single-nucleon potentials is evident from the measured momentum/energy dependence of nucleon optical potential.
It has been shown that the momentum dependence of nuclear mean-field potential affects not only the dynamics of HICs~\cite{Aichelin:1987ti,Gale:1989dm,Zhang:1994hpa,Greco:1999zz,Danielewicz:1999zn,Larionov:2000cu,Persram:2001dg,Chen:2004si,Li:2005jy,Li:2008gp}, but also the thermodynamic properties of nuclear matter~\cite{Csernai:1991fs,Mishra:1993zz,Behera:2001zz,Moustakidis:2007gp,Xu:2007eq,Behera:2011jw,Xu:2014cwa,Constantinou:2015mna,Xu:2019ouo}.

It should be mentioned that for the zero-range Skyrme interaction~\cite{Vautherin:1971aw}, including higher-order derivative terms~\cite{Carlsson:2008gm,Raimondi:2011pz} can introduce additional momentum dependence, which allows for its application in the transport model simulations for HICs, e.g., up to beam energy of about $2$ GeV per nucleon if the N5LO order pseudopotential is used~\cite{Wang:2018yce,Yang:2023umm,Wang:2023zcj,Wang:2024xzq}.
However, the single-nucleon potential in these extended Skyrme interactions will eventually diverge at higher energies due to their polynomial structure, and consequently, the nucleon effective mass in the high-energy limit does not approach the bare mass as it should.
At this point, the finite-range interactions are more realistic for HICs simulations with increasing energy.
In 1980s, the momentum-dependent mean-field potential was first introduced into the BUU model by Gale, Bertsch, and Das Gupta (GBD)~\cite{Gale:1987zz}, and a more realistic parametrization was later proposed by Welke \textit{et al.}~\cite{Welke:1988zz} based on a finite-range force with Yukawa form, which was referred to as the momentum-dependent Yukawa interaction (MDYI).
To incorporate the isospin dependence of the mean-field potential, usually expressed as the Lane potential~\cite{Lane:1962zz}, Bombaci~\cite{Li:2001pet} extended the GBD interaction into the BGBD interaction.
On the other hand, the MDYI interaction was further improved by Das \textit{et al.}~\cite{Das:2002fr}, leading to the so-called MDI interaction, which has been widely used in various transport models (see Refs.~\cite{Li:2008gp,Chen:2013uua} for reviews).

For the finite-range interaction with Yukawa form, a parametrization with two Yukawa couplings was introduced and applied in QMD approach by Maruyama \textit{et al.}~\cite{Maruyama:1997rp}.
Furthermore, the famous Michigan three-range Yukawa (M3Y) interaction was originally established from a bare nucleon-nucleon interaction such as Reid~\cite{Bertsch:1977sg} and Paris~\cite{Anantaraman:1983nxh}, by fitting the Yukawa functions to the G-matrix.
To improve the description of nuclear matter properties with M3Y interaction, DD coupling constants were introduced~\cite{Khoa:1993why,Khoa:1995bzk} and applied to the folding potential for nuclear reactions~\cite{Khoa:1996dzm,Khoa:1997zz}.
On the other side, Nakada introduced zero-range DD terms in the M3Y interaction and developed the M3Y-P$n$ interaction family~\cite{Nakada:2003fw,Nakada:2008eh,Nakada:2009xh,Nakada:2012sq} with constraints from nuclear matter and finite nuclei.
The properties of nuclear matter and neutron stars with these M3Y-type interactions have been investigated in Refs.~\cite{Than:2009ct,Loan:2011aj,Davesne:2025sqs}.

In the present work, we extend the MDI model to a new model called MDI3Y, by including three different MD terms and three zero-range DD terms. The three different MD terms are obtained from the exchange contribution of the central term in the M3Y interaction.
This extension significantly enhances the flexibility in the momentum dependence of both the single-nucleon potential and the symmetry potential, which exhibits several new features compared to the traditional MDI interaction.
Firstly,
the nucleon optical model potential $U_0$ in symmetric nuclear matter at $\rho_0$ can have very different high energy behaviors while nicely describe the empirical nucleon optical potential below about $1$~GeV~\cite{Hama:1990vr,Cooper:1993nx}.
Secondly,
the isoscalar nucleon effective mass $m_{s}^{\ast}$ can have a peak structure at subsaturation densities with a maximum value larger than the bare nucleon mass. Note that the behavior of $m_{s}^{\ast}$ is only determined by fitting the single-nucleon potential to the nucleon optical potential~\cite{Hama:1990vr,Cooper:1993nx}. The $m_{s}^{\ast}/m>1$ at low densities could have significant effects in the mean-field calculations of finite nuclei~\cite{BROWN1963598,Barranco:1980km,Mughabghab:1998zz,Shlomo:1992zcu,Farine:2001oac,Tondeur:2000bd,Goriely:2001zz}.
Thirdly,
the nuclear symmetry potential $U_{\mathrm{sym}}$ can display a non-monotonic momentum dependence by independently adjusting
the linear isospin splitting coefficient $\Delta m_{1}^{\ast}(\rho_0)$ for nucleon effective mass, the isovector nucleon effective mass $m_{v}^{\ast}(\rho_0)$, and the infinite momentum limit value $U_{\mathrm{sym}}^{\infty}$ of the $U_{\mathrm{sym}}$ at $\rho_0$.
The non-monotonic $U_{\mathrm{sym}}$ implies that the neutron-proton effective mass splitting $m_{\rm n\text{-}p}^{\ast}$ may have different signs at different energies, which could be tested in transport model simulations for HICs by investigating the light particle emission~\cite{Li:2005by,Coupland:2014gya,Zhang:2014sva,Kong:2015rla,Wang:2023buv}.

In addition, in the MDI3Y model, the many-body force is mimicked by three zero-range DD terms, following the idea of Fermi momentum expansion~\cite{Patra:2022yqc}.
Consequently, the density behaviors of the SNM EOS and $E_{\mathrm{sym}}$ become highly flexible, allowing for reasonable descriptions of neutron star properties.
We notice that when specific parameters are selected, the DD term in the MDI3Y model will reduce to the form in M3Y-P$n$ model~\cite{Nakada:2003fw,Nakada:2008eh,Nakada:2009xh,Nakada:2012sq}.
Based on the MDI3Y model, we construct interactions with $L=35$, $55$, and $75$~MeV for the density slope parameter of the symmetry energy.
Moreover, for the case of $L=35$~MeV, we also employ four different $U_{\mathrm{sym}}$, including both monotonic and non-monotonic types.
In future, these new interactions can be used in transport model simulations for HICs at intermediate and high energies, through which one may obtain a more comprehensive understanding of SNM EOS and $E_{\mathrm{sym}}$ as well as $U_0$, $U_{\mathrm{sym}}$ and $m_{\rm n\text{-}p}^{\ast}$.

This paper is organized as follow:
In Sec.~\ref{sec:framework},
we introduce the theoretical framework of the MDI3Y model, and we present the expressions of energy density and single-nucleon potential under general nonequilibrium conditions.
In Sec.~\ref{sec:fitting},
we discuss the fitting procedure to construct MDI3Y interactions, along with the data/constraints used for fitting.
In Sec.~\ref{sec:properties},
we present several properties for the cold nuclear matter, including the bulk properties, the single-nucleon potential, the first-order and second-order symmetry potentials, and the nucleon effective masses. The neutron star properties for the MDI3Y interactions are also presented.
In Sec.~\ref{sec:summary},
we summarize this work and make a brief outlook.
In Appendix~\ref{sec:Nakada},
we present the relations between the parameters in the MDI3Y interaction and M3Y-P$n$ interaction.
In Appendix~\ref{sec:HF_DME},
we present some details of the derivations for the energy density using Hartree-Fock approximation, and we also discuss the density matrix expansion for the MDI3Y interaction.
For completeness, we provide the expressions related to the main text in Appendix~\ref{sec:Apd_1}.

\section{Theoretical framework}
\label{sec:framework}
\subsection{The MDI3Y interactions}
We assume that the central part $V^{C}$ of the interaction between two nucleons located at $\vec{r}_{1}$ and $\vec{r}_{2}$ is composed of three Yukawa interactions:
\begin{equation}
\small
\label{eq:v3Y}
V^{C}\left(\vec{r}_1, \vec{r}_2\right)=
\sum_{i=1}^{3}  \left(W_{i}+B_{i} P_\sigma-H_{i} P_\tau-M_{i} P_\sigma P_\tau\right) \frac{e^{-\mu_{i}\left|\vec{r}_1-\vec{r}_2\right|}}{\mu_{i}\left|\vec{r}_1-\vec{r}_2\right|},
\end{equation}
where $P_\sigma$ and $P_\tau$ are the spin and isospin exchange operators, respectively.
Following the idea of Fermi momentum expansion~\cite{Patra:2022yqc,Wang:2023zcj}, three zero-range DD terms are introduced to mimic the many-body force as follows:
\begin{equation}
\label{eq:V_DD}
V^{\mathrm{DD}}= \sum_{n=1,3,5} \frac{1}{6} t_{3}^{\left[ n \right] }\left(1+x_{3}^{\left[ n \right] } P_\sigma\right) \rho^{n/3}(\vec{r}) \delta(\vec{r}_{1}-\vec{r}_{2}),
\end{equation}
where $\vec{r}=(\vec{r}_{1}+\vec{r}_{2})/2$.
As a result, the EOS of nuclear matter can be expressed as a power series in $\rho^{1/3}$ [see Eq.~(\ref{eq:EOS3Y}) and the relevant discussion in Sec.~\ref{sec:cold_EOS}].
It should be noted that including additional terms with larger $n$ in Eq.~(\ref{eq:V_DD}) can provide more flexible high-density behavior of nuclear matter EOS, although the DD term up to $n=5$ is sufficient to satisfy the current constraints.

In general, the complete nuclear effective interaction also includes spin-orbit and tensor components, which are important for nuclear structure calculations~\cite{Nakada:2001nv,Nakada:2005hm,Nakada:2008kq}.
In the present work, we focus on the spin-averaged quantities, where the spin-dependent terms make no contribution.
Therefore, the MDI3Y interaction used in this work is written as
\begin{equation}
\label{eq:v_total}
v_{\mathrm{MDI3Y}} = V^{C} + V^{\mathrm{DD}}.
\end{equation}
It should be noted that Eqs.~(\ref{eq:v3Y}) and (\ref{eq:V_DD}) can be equivalently expressed as the combinations of different spin-isospin (S-T) channels, as the forms in the M3Y-P$n$ interactions~\cite{Nakada:2003fw,Nakada:2008eh,Nakada:2009xh,Nakada:2012sq}.
In Appendix~\ref{sec:Nakada}, we provide the relations between the parameters for these two representations, which could be useful for future investigations about the S-T structures and properties of finite nuclei.

\subsection{Hamiltonian density and single-nucleon potential in one-body transport model}
The total energy of the nuclear system can be obtained from
\begin{equation}
\label{eq:E_pot}
\begin{aligned}
\mathrm{H} = & \sum_{i} \langle i| \frac{p^2}{2m} |i \rangle  +\frac{1}{2} \sum_{i,j} \langle i j| v_{\mathrm{MDI3Y}}(1-P_M P_\sigma P_\tau) |ij\rangle\\
=& \int \mathcal{H}(\vec{r})  \mathrm{d}^3 r ,
\end{aligned}
\end{equation}
where $P_M$ is the space exchange operator.
The $m$ is the average nucleon mass in free-space throughout this paper.

During the HICs, the nucleons are generally far from thermal equilibrium, and in transport models they are described by the phase-space distribution function (Wigner function) $f_\tau \left( \vec{r}, \vec{p}\right)$, with $\tau=1$ (or n) for neutrons and $-1$ (or p) for protons.
The $f_\tau \left( \vec{r}, \vec{p}\right)$ becomes the Fermi-Dirac distribution in momentum space when the system approaches to equilibrium.
Based on Eq.~(\ref{eq:E_pot}), the total Hamiltonian density $\mathcal{H}(\vec{r})$ of the collision system can be expressed as (see Appendix~\ref{sec:Vc_HF} for details)
\begin{equation}
\mathcal{H}(\vec{r}) = \mathcal{H}^{\mathrm{kin}}(\vec{r})
+ \mathcal{H}^{\mathrm{loc}}(\vec{r})
+ \mathcal{H}^{\mathrm{MD}}(\vec{r})
+ \mathcal{H}^{\mathrm{DD}}(\vec{r}),
\end{equation}
where $\mathcal{H}^{\mathrm{kin}}(\vec{r})$, $\mathcal{H}^{\mathrm{loc}}(\vec{r})$, $\mathcal{H}^{\mathrm{MD}}(\vec{r})$, and $\mathcal{H}^{\mathrm{DD}}(\vec{r})$ are, respectively, the kinetic, local, MD, and DD terms.
The kinetic term can be expressed as
\begin{equation}
\label{eq:Hkin}
\mathcal{H} ^{ \mathrm{kin}  } \left( \vec{r} \right)
=\sum_{\tau=\mathrm{n},\mathrm{p}} \int {\rm d}^3 p\frac{p^2}{2 m} f_\tau \left( \vec{r}, \vec{p} \right).
\end{equation}

The local and MD terms come from the direct and exchange terms of the finite-range Yukawa interactions.
Within Hartree-Fock approximation, the local term can be expressed as
\begin{equation}
\label{eq:Hdr_MainTXT}
\begin{aligned}
\mathcal{H}^{\mathrm{loc}}(\vec{r})
=& \frac{1}{2} \sum_{i=1}^{3} \int \mathrm{d}^3 s
\frac{e^{-\mu_i s}}{\mu_i s}\bigg\{ (W_i + \frac{B_i}{2}-H_i-\frac{M_i}{2}) \\
&\times  \Big[ \rho_{\mathrm{n}}(\vec{r}+ \frac{\vec{s}}{2}) \rho_{\mathrm{n}}(\vec{r}- \frac{\vec{s}}{2}) + \rho_{\mathrm{p}}(\vec{r}+ \frac{\vec{s}}{2}) \rho_{\mathrm{p}}(\vec{r}- \frac{\vec{s}}{2})\Big] \\
&+ 2 (W_i+\frac{B_i}{2}) \rho_{\mathrm{n}}(\vec{r}+ \frac{\vec{s}}{2}) \rho_{\mathrm{p}}(\vec{r}- \frac{\vec{s}}{2})
\bigg\},
\end{aligned}
\end{equation}
where $\vec{s}=(\vec{r}_{1}-\vec{r}_{2})$. Under the approximation of uniform nuclear matter with constant density by assuming the densities vary slowly in space, Eq.~(\ref{eq:Hdr_MainTXT}) can be obtained approximately as
\begin{equation}
\label{eq:H_loc}
\mathcal{H}^{\mathrm{loc}}(\vec{r}) = \frac{A_l}{2 \rho_0}\left(\rho_{\mathrm{n}}^2+\rho_{\mathrm{p}}^2\right)
+ \frac{A_u \rho_{\mathrm{n}} \rho_{\mathrm{p}}}{\rho_0},
\end{equation}
which is exactly equivalent to the local term in the original MDI model~\cite{Das:2002fr,Chen:2013uua} and will be also adopted in the following for the MDI3Y model calculations.
Furthermore, the MD term can be expressed as
\begin{equation}
\label{eq:H_MD}
\mathcal{H}^{\mathrm{MD}}(\vec{r}) = \frac{1}{\rho_0} \sum_{i=1}^{3} \sum_{\tau, \tau^{\prime}} C_{\tau \tau^{\prime},i}
\iint \mathrm{d}^3 p \mathrm{~d}^3 p^{\prime} \frac{f_\tau(\vec{r}, \vec{p}) f_{\tau^{\prime}}\left(\vec{r}, \vec{p}^{\, \prime}\right)}{1+\left(\vec{p}-\vec{p}^{\, \prime}\right)^2 / \Lambda_{i}^2}.
\end{equation}

Based on Eq.~(\ref{eq:V_DD}), the DD term can be expressed as
\begin{equation}
\label{eq:H_DD}
\begin{aligned}
\mathcal{H}^{\mathrm{DD}}(\vec{r}) = &
\sum_{n=1,3,5} \frac{1}{24} t_3^{[n]} \left[ \left(2+x_3^{[n]}\right) \rho^2 \right. \\
& \left. -\left(2 x_3^{[n]}+1\right) ( \rho_{\mathrm{n}}^2 + \rho_{\mathrm{p}}^2 )\right] \rho^{n / 3},
\end{aligned}
\end{equation}
where $\rho=\rho_{\mathrm{n}} + \rho_{\mathrm{p}}$.

Comparing Eqs.~(\ref{eq:H_loc}) and (\ref{eq:H_MD}) with Eqs.~(\ref{eq:Hdr_general_NM}) and (\ref{eq:Her_f}) in Appendix, we can obtain the relations among the parameters in the energy density of uniform nuclear matter and those in the two-body interaction:
\begin{align}
\label{eq:Lmbd}
\Lambda_{i} &= \hbar \mu_{i}~~(i=1,2,3),\\
\label{eq:dir1}
\frac{1}{\rho_0} A_{l} &= \sum_{i=1}^{3} \left[
\frac{4\pi}{\mu_{i}^{3}}\left( W_{i} + \frac{B_i}{2} - H_i - \frac{M_i}{2} \right)\right], \\
\label{eq:dir2}
\frac{1}{\rho_0}A_{u} &= \sum_{i=1}^{3} \left[
\frac{4\pi}{\mu_{i}^{3}}\left( W_{i} + \frac{B_i}{2}
\right)\right], \\
\label{eq:exch1}
\frac{1}{\rho_0} C_{l,i} &= \frac{2\pi}{\mu_{i}^{3}} \left(
M_i + \frac{H_i}{2} -B_i - \frac{W_i}{2} \right)~~(i=1,2,3), \\
\label{eq:exch2}
\frac{1}{\rho_0} C_{u,i} &= \frac{2\pi}{\mu_{i}^{3}} \left( M_i +\frac{H_i}{2} \right)~~(i=1,2,3),
\end{align}
where $C_{l,i} \equiv C_{\tau, \tau,i}$ and $C_{u,i} \equiv C_{\tau, -\tau,i}$.
The force-ranges $\mu_i^{-1}$ correspond to the Compton wavelengths of the mesons with masses of $\Lambda_i c$.
The $A_l$ and $A_u$ correspond to the direct term, while $C_{l,i}$ and $C_{u,i}$ describe the exchange term.
It can be seen from Eq.~(\ref{eq:Hdr_general_NM}) that the direct contributions from different Yukawa interactions are degenerate for uniform nuclear matter.
This leads to the fact that in the MDI3Y model for spin-saturated nuclear matter considered in the present work, there are $17$ model parameters, namely,
$A_l$, $A_u$; $C_{l,i}$, $C_{u,i}$, $\Lambda_i$ ($i=1,2,3$); $t_{3}^{[n]}$, $x_{3}^{[n]}$ ($n=1,3,5$), while the number of parameters in Eqs.~(\ref{eq:v3Y}) and (\ref{eq:V_DD}) is $21$.
To fully determine the parameters for the two-body interaction, one can impose some additional constraints, e.g., the S-T structure of nuclear matter energy density, which can be obtained from microscopic calculations for spin-unsaturated nuclear matter~\cite{Baldo:1997ag,Baldo:2014yja,Davesne:2016fqg,Batail:2022bjq}.
It is interesting to note that if the two-body interaction is taken as one Yukawa term, its parameters can be uniquely determined by the corresponding MDI interaction~\cite{Xu:2010ce}.

It should be pointed out that Eqs.~(\ref{eq:Hkin}), (\ref{eq:H_loc}), (\ref{eq:H_MD}) and (\ref{eq:H_DD}) fully express the Hamiltonian density for nucleonic system under general nonequilibrium condition for the MDI3Y model. Furthermore, the Hamiltonian density of the MDI3Y model can be reduced to the original Hamiltonian density of the MDI model by using one Yukawa term contribution in Eq.~(\ref{eq:H_MD}) and changing the DD contribution [i.e., Eq.~(\ref{eq:H_DD})] to the original DD term in the MDI model. In addition, one can see that the Hamiltonian density in the MDI3Y model can be obtained from Hartree-Fock calculation by uniform nuclear matter approximation for the direct term contribution of the finite-range interaction.

\begin{widetext}
Based on the Landau Fermi liquid theory, the single-nucleon potential can be calculated as
\begin{equation}
\label{eq:U=dHdf}
U_{\tau} \left( \vec{r}, \vec{p} \right)
= \frac{\delta \mathrm{H}^{\mathrm{pot}}}{ \delta f_\tau\left( \vec{r}, \vec{p} \right)}
= \frac{ \partial \left[ \mathcal{H}^{\mathrm{loc}} \left( \vec{r} \right) + \mathcal{H}^{\mathrm{DD}} \left( \vec{r} \right)  \right] }
{\partial \rho_\tau(\vec{r})}
+ \frac{\delta \mathrm{H}^{\mathrm{MD}}}{ \delta f_\tau\left( \vec{r}, \vec{p} \right)},
\end{equation}
where $\mathrm{H}^{\mathrm{pot}}= \int \mathrm{d} \vec{r} \left[ \mathcal{H}^{\mathrm{loc}}(\vec{r})+\mathcal{H}^{\mathrm{DD}}(\vec{r})+\mathcal{H}^{\mathrm{MD}}(\vec{r}) \right]$ is the potential part of the Hamiltonian with $\mathrm{H}^{\mathrm{MD}}= \int \mathrm{d} \vec{r} \mathcal{H}^{\mathrm{MD}}(\vec{r}) $ being the MD part.
Substituting Eqs.~(\ref{eq:H_loc})-(\ref{eq:H_DD}) into Eq.~(\ref{eq:U=dHdf}), we then obtain the single-nucleon potential in nuclear matter with the MDI3Y interactions as
\begin{equation}
\label{eq:Utau}
\begin{aligned}
U_{\tau}(\vec{r}, \vec{p})= & A_l \frac{\rho_\tau}{\rho_0} + A_u \frac{\rho_{-\tau}}{\rho_0}  + \sum_{n=1,3,5} \left\{ \frac{1}{12} t_{3} ^{\left[n \right]}\left[\left(2+x_{3} ^{\left[n \right]}\right) \rho(\vec{r})-\left(2 x_{3} ^{\left[n \right]}+1\right) \rho_\tau(\vec{r})\right] \rho(\vec{r})^{\frac{n}{3}}  \right\}
\\
&+\sum_{n=1,3,5} \left\{ \frac{t_{3} ^{\left[n \right]}}{24} \frac{n}{3} \left[\left(2+x_{3} ^{\left[n \right]}\right) \rho(\vec{r})^2-\left(2 x_{3} ^{\left[n \right]}+1\right) \sum_{\tau=\mathrm{n}, \mathrm{p}} \rho_\tau(\vec{r})^2\right] \rho(\vec{r})^{ \frac{n}{3}-1 }  \right\}  \\
& + \frac{2}{\rho_0} \sum_{i=1}^{3} \int \mathrm{d}^3 p^{\prime} \frac{ C_{l,i} ~f_\tau(\vec{r}, \vec{p}^{\, \prime})+ C_{u,i} ~ f_{-\tau}\left(\vec{r}, \vec{p}^{\, \prime}\right)}{1+\left(\vec{p}-\vec{p}^{\, \prime}\right)^2 / \Lambda_{i}^2}.
\end{aligned}
\end{equation}

In Appendix~\ref{sec:DME}, we discuss the density matrix expansion for the MDI3Y interaction, with the exchange contribution from finite-range interaction being approximated by a Skyrme-like zero-range interaction with density-dependent parameters.
The resulting energy density functional and single-nucleon potential can be expressed in terms of local densities and their spatial gradients, which could be used to study finite nuclei properties~\cite{Hofmann:1997zu}, transition density of neutron star~\cite{Xu:2010ce,Xu:2009vi}, as well as in transport models~\cite{Wang:2019ghr,Wang:2020ixf} to self-consistently obtain the initial states of nuclei and perform Lattice BUU simulations for HICs.

\subsection{Equation of state of cold nuclear matter}
\label{sec:cold_EOS}
The EOS of isospin asymmetric nuclear matter with total nucleon density $\rho=\rho_{\mathrm{n}} + \rho_{\mathrm{p}}$ and isospin asymmetry $\delta = (\rho_{\mathrm{n}}-\rho_{\mathrm{p}})/\rho$ are defined as its binding energy per nucleon.
For the uniform nuclear matter at zero temperature, $f_{\tau} ( \vec{r},\vec{p} )$ becomes a step function in momentum space, i.e., $f_{\tau} ( \vec{r},\vec{p} )=g_{\tau} \theta(p_{F_{\tau}}-|\vec{p}|)$, where $g_{\tau}$ is the degeneracy and $p_{F_{\tau}}=\hbar(3 \pi^2 \rho_\tau)^{1/3}$ is the Fermi momentum of nucleons with isospin $\tau$.
The EOS of cold nuclear matter can then be analytically expressed as
\begin{equation}
\label{eq:EOS3Y}
\begin{aligned}
E(\rho,\delta)= & \frac{3}{5} \frac{a^2 \hbar^2}{2m} F_{5/3} \rho^{2/3}
+ \frac{1}{4\rho_0}\left[ A_l(1+\delta^2) + A_u(1-\delta^2)  \right] \rho^{3/3} \\
&+ \sum_{i=1}^{3}  \sum_{\tau, \tau^{\prime}} C_{\tau \tau^{\prime},i} \, E^{\mathrm{MD}}_{\tau \tau^{\prime},i}
+\sum_{n=1,3,5} \frac{1}{48} t_{3}^{[n]} \left[ 3-(2x_{3}^{[n]} +1) \delta^2 \right] \rho^{\frac{n}{3} +1},
\end{aligned}
\end{equation}
where $a=(3\pi^2/2)^{1/3}$ and $F_x =\left[(1+\delta)^x+(1-\delta)^x\right] /2$.
In Eq.~(\ref{eq:EOS3Y}), we have defined the contribution of the MD term as~\cite{Chen:2007ih}
\begin{equation}
\label{eq:E_MD}
\begin{aligned}
E^{\mathrm{MD}}_{\tau \tau^{\prime},i} =& \frac{\Lambda_{i}^{2}}{24 \pi^{4} \hbar^{6} \rho_{0} \rho} \Bigg\{ p_{F,\tau} p_{F,\tau^{\prime}} \Big[3 (p_{F,\tau}^{2} + p_{F,\tau^{\prime}}^{2}) -\Lambda_{i}^{2} \Big] + 4 \Lambda_{i} \bigg[ (p_{F,\tau}^{3} - p_{F,\tau^{\prime}}^{3}) \arctan \frac{p_{F,\tau} - p_{F,\tau^{\prime}}}{\Lambda_{i}} \\
&- (p_{F,\tau}^{3} + p_{F,\tau^{\prime}}^{3}) \arctan \frac{p_{F,\tau} + p_{F,\tau^{\prime}}}{\Lambda_{i}} \bigg]
+\frac{1}{4} \bigg[ \Lambda_{i}^{4} + 6\Lambda_{i}^2 (p_{F,\tau}^{2} + p_{F,\tau^{\prime}}^{2}) -3 (p_{F,\tau}^{2} - p_{F,\tau^{\prime}}^{2})^2
\bigg] \ln \frac{(p_{F,\tau} + p_{F,\tau^{\prime}})^2 + \Lambda_{i}^2}{(p_{F,\tau} - p_{F,\tau^{\prime}})^2 + \Lambda_{i}^2}
\Bigg\}.
\end{aligned}
\end{equation}
The $E^{\mathrm{MD}}_{\tau \tau^{\prime}}$ can be expanded as a power series of $p_{F_{\tau}}$ (here we consider the case of SNM with $p_{F_{\tau}}=p_{F}=a \hbar \rho^{1/3}$):
\begin{equation}
\label{eq:EOS_Fermi}
E^{\mathrm{MD}}_{\tau \tau^{\prime}} \propto \frac{8}{3} p_{F}^{3}
- \frac{16 p_{F}^{5}}{5 \Lambda^{2}}
+ \frac{192 p_{F}^{7}}{35 \Lambda^{4}}
- \frac{512 p_{F}^{9}}{45 \Lambda^{6}} + \mathcal{O}(p_{F}^{11}).
\end{equation}
It can been seen from Eq.~(\ref{eq:EOS3Y}) that the inclusion of the DD term as in Eq.~(\ref{eq:H_DD}) leads to the EOS that can be expressed as a power series in $\rho^{1/3}$ (equivalently the Fermi momentum $p_F$), from $\rho^{2/3}$ to $\rho^{9/3}$, while the contributions from DD terms and MD terms can be clearly distinguished.
We would like to mention that different choices of power exponent in DD term and their impact on the nuclear EOS have been investigated in Refs.~\cite{Yang:2016mvq,Burrello:2020myg}.
By employing different powers for $p_F$ in DD term compared to MD term, the corresponding nuclear EOS may better reproduce the low-density behaviors predicted by microscopic calculations~\cite{Yang:2016nkd,Wellenhofer:2021eis}.
\end{widetext}

The EOS of isospin asymmetric nuclear matter can be expanded as a power series in $\delta$, i.e.,
\begin{equation}
E\left(\rho, \delta\right) = E_{0}\left(\rho\right) + E_{\mathrm{sym}}\left(\rho\right)\delta^2 + E_{\mathrm{sym},4}\left(\rho\right)\delta^4 + \mathcal{O}\left(\delta^6\right),
\end{equation}
where $E_0(\rho)$ is the EOS of the SNM.
The symmetry energy $E_{\mathrm{sym}}(\rho)$ and the fourth-order symmetry energy $E_{\mathrm{sym},4}(\rho)$ can be expressed as
\begin{equation}
\label{eq:Esym_def}
E_{\mathrm{sym}}(\rho)  =\left.\frac{1}{2 !} \frac{\partial^2 E(\rho, \delta)}{\partial \delta^2}\right|_{\delta=0},
\end{equation}
and
\begin{equation}
\label{eq:Esym4_def}
E_{\mathrm{sym}, 4}(\rho)  =\left.\frac{1}{4 !} \frac{\partial^4 E(\rho, \delta)}{\partial \delta^4}\right|_{\delta=0},
\end{equation}
respectively.
The expressions of $E_{0}(\rho)$, $E_{\mathrm{sym}}(\rho)$, and $E_{\mathrm{sym}, 4}(\rho)$ are presented in Appendix~\ref{sec:Apd_1}.

The pressure of the isospin asymmetric nuclear matter can be expressed as
\begin{equation}
\label{eq:press}
P \left(\rho,\delta \right) = \rho^2 \frac{\partial E(\rho, \delta)}{\partial \rho},
\end{equation}
and the saturation density $\rho_{0}$ is defined as the density where the pressure of SNM is zero (except for $\rho=0$), i.e.,
\begin{equation}
\label{eq:rho_0}
\left. P \left(\rho_0, \delta=0 \right) = \rho_{0}^2 \frac{\mathrm{d} E(\rho, 0)}{\mathrm{d} \rho} \right| _{\rho=\rho_0}=0 .
\end{equation}
Around $\rho_{0}$, both $E_0(\rho)$ and $E_{\mathrm{sym}}(\rho)$ can be expanded as a power series in a dimensionless variable $\chi \equiv \frac{\rho-\rho_0}{3 \rho_0}$~\cite{Chen:2009wv}, i.e.,
\begin{equation}
\label{eq:E0_taylor}
\begin{aligned}
E_0(\rho)=&\,E_0\left(\rho_0\right)+L_0 \chi+\frac{K_0}{2 !} \chi^2\\
&+\frac{J_0}{3 !} \chi^3 +\frac{I_0}{4 !} \chi^4+ \frac{H_0}{5 !} \chi^5 +\mathcal{O}\left(\chi^6\right),
\end{aligned}
\end{equation}
and
\begin{equation}
\label{eq:Esym_taylor}
\begin{aligned}
E_{\mathrm{sym}}(\rho)= & \, E_{\mathrm{sym}}\left(\rho_0\right)+L \chi+\frac{K_{\mathrm{sym}}}{2 !} \chi^2 \\
&+\frac{J_{\mathrm{sym}}}{3 !} \chi^3
+\frac{I_{\mathrm{sym}}}{4 !} \chi^4 +\frac{H_{\mathrm{sym}}}{5 !} \chi^5
+\mathcal{O}\left(\chi^6\right).
\end{aligned}
\end{equation}
The coefficients of $\chi^n$ in Eqs.~(\ref{eq:E0_taylor}) and (\ref{eq:Esym_taylor}) are commonly used to characterize the density dependence of $E_0(\rho)$ and $E_{\mathrm{sym}}(\rho)$, respectively, which can be expressed as
\begin{align}
\label{eq:L_def}
L_0 &  =\left.3 \rho_0 \frac{ \mathrm{d} E_0(\rho)}{\mathrm{d} \rho}\right|_{\rho=\rho_0},
L  =\left.3 \rho_0 \frac{\mathrm{d} E_{\mathrm{sym}}(\rho)}{\mathrm{d}\rho}\right|_{\rho=\rho_0},\\
\label{eq:K_def}
K_0 &  =\left. (3\rho_0)^2 \frac{\mathrm{d}^2 E_0(\rho)}{\mathrm{d} \rho^2}\right|_{\rho=\rho_0},
K_{\mathrm{sym}}  =\left.(3\rho_0)^2 \frac{\mathrm{d}^2 E_{\mathrm{sym}}(\rho)}{\mathrm{d} \rho^2}\right|_{\rho=\rho_0},\\
\label{eq:J_def}
J_0 &  =\left.(3\rho_0)^3 \frac{\mathrm{d}^3 E_0(\rho)}{\mathrm{d} \rho^3}\right|_{\rho=\rho_0},
J_{\mathrm{sym}}=\left.(3\rho_0)^3 \frac{\mathrm{d}^3 E_{\mathrm{sym}}(\rho)}{\mathrm{d} \rho^3}\right|_{\rho=\rho_0},\\
\label{eq:I_def}
I_0 & =\left.(3\rho_0)^4 \frac{\mathrm{d}^4 E_0(\rho)}{\mathrm{d} \rho^4}\right|_{\rho=\rho_0},
I_{\mathrm{sym}}=\left.(3\rho_0)^4 \frac{\mathrm{d}^4 E_{\mathrm{sym}}(\rho)}{\mathrm{d} \rho^4}\right|_{\rho=\rho_0},\\
\label{eq:H_def}
H_0 & =\left.(3\rho_0)^5 \frac{\mathrm{d}^5 E_0(\rho)}{\mathrm{d} \rho^5}\right|_{\rho=\rho_0},
H_{\mathrm{sym}}=\left.(3\rho_0)^5 \frac{\mathrm{d}^5 E_{\mathrm{sym}}(\rho)}{\mathrm{d} \rho^5}\right|_{\rho=\rho_0}.
\end{align}
Obviously, we have $L_0=0$ by the definition of $\rho_0$ in Eq.~(\ref{eq:rho_0}).
$K_0$ is known as the incompressibility coefficient of SNM which characterizes the curvature of $E_0(\rho)$ at $\rho_0$.
The higher-order coefficients $J_0$, $I_0$ and $H_0$ are commonly referred to as the skewness, kurtosis, and hyper-skewness coefficients of SNM.
Correspondingly, $L$, $K_{\mathrm{sym}}$, $J_{\mathrm{sym}}$, $I_{\mathrm{sym}}$, and $H_{\mathrm{sym}}$ are the slope, curvature, skewness, kurtosis, and hyper-skewness coefficient of $E_{\mathrm{sym}}(\rho)$ at $\rho_0$, respectively.
The expressions of these characteristic parameters are presented in Appendix~\ref{sec:Apd_1}.

\begin{widetext}
\subsection{Single-nucleon potential, symmetry potential, nucleon effective masses and its linear isospin splitting coefficient in cold nuclear matter}
In the cold nuclear matter, the single-nucleon potential [Eq.~(\ref{eq:Utau})] can be analytically expressed as a function of $\rho$, $\delta$, and the magnitude of nucleon momentum $p=\left|\vec{p}\right|$, i.e.,
\begin{equation}
\label{eq:Utau_T0}
\begin{aligned}
U_{\tau}(\rho,\delta,p)=& A_l \frac{\rho_{\tau}}{\rho_0}
+ A_u \frac{\rho_{-\tau}}{\rho_0} +  \sum_{i=1}^{3}  \sum_{\tau^{\prime}} C_{\tau \tau^{\prime},i} \, U^{\mathrm{MD}}_{\tau^{\prime},i} +\sum_{n=1,3,5} \frac{1}{24} t_3^{[n]} \left[ (3+\frac{n}{2}) - (1+2 x_3^{[n]}) ( \tau \delta + \frac{n}{6} \delta^2 ) \right] \rho^{\frac{n}{3} +1} ,
\end{aligned}
\end{equation}
where~\cite{Chen:2007ih}
\begin{equation}
\label{eq:UtauMD_T0}
\begin{aligned}
U^{\mathrm{MD}}_{\tau^{\prime},i} = \frac{\Lambda_{i}^{3}}{2\pi^2 \hbar^3 \rho_0} \Bigg[ \frac{p_{F,\tau^{\prime}}^{2} +\Lambda_{i}^{2} -p^2}{2 p \Lambda_{i}}  \ln \frac{(p + p_{F,\tau^{\prime}})^2 + \Lambda_{i}^2}{(p - p_{F,\tau^{\prime}})^2 + \Lambda_{i}^2}
+ \frac{2 p_{F,\tau^{\prime}}}{\Lambda_{i}} - 2 \arctan \frac{p + p_{F,\tau^{\prime}}}{\Lambda_{i}}
+ 2\arctan \frac{p - p_{F,\tau^{\prime}}}{\Lambda_{i}} \Bigg].
\end{aligned}
\end{equation}

The $U_\tau(\rho, \delta,p)$ can be expanded as a power series in $\tau\delta$, i.e.,
\begin{equation}
\label{eq:Utau_series}
U_\tau(\rho, \delta, p) = U_0(\rho,p)+\sum_{i=1,2, \cdots} U_{\mathrm{sym}, i}(\rho,p)(\tau \delta)^i
= U_0(\rho,p)+U_{\mathrm{sym}, 1}(\rho,p)(\tau \delta)+U_{\mathrm{sym}, 2}(\rho, p)(\tau \delta)^2+\cdots,
\end{equation}
where
\begin{equation}
U_0(\rho,p)=U_{\tau}(\rho,\delta=0,p)
\end{equation}
is the single-nucleon potential in SNM, and
\begin{equation}
U_{\mathrm{sym}, i}(\rho,p) \left.\equiv \frac{1}{i !} \frac{\partial^i U_n(\rho, \delta,p)}
{\partial \delta^i}\right|_{\delta=0} =\left.\frac{(-1)^i}{i !} \frac{\partial^i U_p(\rho, \delta,p)}{\partial \delta^i}\right|_{\delta=0}.
\end{equation}
is the $i$th-order symmetry potential.
Neglecting higher-order terms ($\delta^2,\delta^3,\cdots$) in Eq.~(\ref{eq:Utau_series}) leads to the well-known Lane potential~\cite{Lane:1962zz}:
\begin{equation}
\label{eq:Lane}
U_\tau(\rho, \delta, p) \approx U_0(\rho,p)+U_{\mathrm{sym }}(\rho,p)(\tau \delta).
\end{equation}
In the MDI3Y interaction, $U_0(\rho,p)$ can be expressed as
\begin{equation}
\label{eq:U0_C}
U_0(\rho,p) = C_0(\rho) + C_{1} f_{\Lambda_{1}}(\rho,p)
+ C_{2} f_{\Lambda_{2}}(\rho,p) + C_{3} f_{\Lambda_{3}}(\rho,p),
\end{equation}
where
\begin{equation}
\label{eq:flambda}
f_{\Lambda_i}(p) = \frac{3 \Lambda^3_i}{4 p_{F_0}^{3}} \bigg[
\frac{p_{F}^2 + \Lambda^2_i -p^2}{2 p \Lambda_i} \ln \frac{(p+p_{F})^{2}+\Lambda^2_i}{(p-p_{F})^{2}+\Lambda^2_i} + \frac{2 p_{F}}{\Lambda_i} - 2\arctan \frac{p+p_{F}}{\Lambda_i} +2\arctan \frac{p-p_{F}}{\Lambda_i} \bigg],
\end{equation}
\begin{equation}
\label{eq:C0}
C_0(\rho) = U_{0}(\rho,p=\infty) = \frac{A_l+A_u}{2} \frac{\rho}{\rho_0} + \sum_{n=1,3,5} \frac{t_{3}^{[n]}}{16}\left( \frac{n}{3} +2 \right) \rho^{\frac{n}{3}+1},
\end{equation}
\begin{equation}
\label{eq:Ci}
C_{i} = C_{l,i} + C_{u,i} ~~ (i=1,2,3),
\end{equation}
and $p_{F_0}=a \hbar \rho_0^{1/3}$ is the Fermi momentum of nucleons in the SNM at $\rho_0$.
The (first-order) symmetry potential can be expressed as
\begin{equation}
\label{eq:Usym_D}
U_{\mathrm{sym}}(\rho,p) = D_0(\rho) + D_{1} g_{\Lambda_{1}}(\rho,p)
+ D_{2} g_{\Lambda_{2}}(\rho,p) + D_{3} g_{\Lambda_{3}}(\rho,p),
\end{equation}
where
\begin{equation}
\label{eq:gLambda}
g_{\Lambda_i}(\rho,p) = \frac{1}{4} \frac{p_{F}^{2} \Lambda^2_i}{p_{F_0}^{3} \, p} \ln \frac{(p+p_{F})^{2}+\Lambda^2_i}{(p-p_{F})^{2}+\Lambda^2_i},
\end{equation}
\begin{equation}
\label{eq:D0}
D_0(\rho) = U_{\mathrm{sym}}(\rho,p=\infty) = \frac{A_l-A_u}{2} \frac{\rho}{\rho_0} -\sum_{n=1,3,5} \frac{1}{24} t_3^{[n]}\left(2 x_3^{[n]}+1\right) \rho^{\frac{n}{3}+1},
\end{equation}
and
\begin{equation}
\label{eq:Di}
D_{i} = C_{l,i} - C_{u,i} ~~ (i=1,2,3).
\end{equation}
The second-order symmetry potential can be expressed as
\begin{equation}
\label{eq:Usym2}
U_{\mathrm{sym},2}(\rho,p) = - \sum_{n=1,3,5} \frac{t_{3}^{[n]}}{48} \frac{n}{3}
\left( 2 x_{3}^{[n]} +1  \right) \rho^{ \frac{n}{3} +1}
+ C_{1} h_{\Lambda_1}(\rho,p)
+ C_{2} h_{\Lambda_2}(\rho,p)
+ C_{3} h_{\Lambda_3}(\rho,p),
\end{equation}
with $C_i$ defined in Eq.~(\ref{eq:Ci}) and
\begin{equation}
\label{eq:hLambda}
h_{\Lambda_i}(\rho,p) = \frac{1}{6} \frac{\rho}{\rho_0} \frac{\Lambda_i^2 \left( p^2 + \Lambda_i^2 - p_F^2 \right)}{\left[ (p+p_{F})^{2}+\Lambda^2_i \right] \left[ (p-p_{F})^{2}+\Lambda^2_i \right]}
- \frac{1}{24}  \frac{p_{F}^{2} \Lambda^2_i}{p_{F_0}^{3} \, p} \ln \frac{(p+p_{F})^{2}+\Lambda^2_i}{(p-p_{F})^{2}+\Lambda^2_i}.
\end{equation}

The momentum dependence of the single-nucleon potential is characterized by the nucleon effective mass~\cite{Jaminon:1989wj,Li:2018lpy}, and it can be expressed as
\begin{equation}
\label{eq:mtau_def}
\frac{m_{\tau}^{\ast}(\rho,\delta)}{m}=\left[1+\left.\frac{m}{p} \frac{\mathrm{d} U_\tau(\rho, \delta,p)}{\mathrm{d} p}\right|_{p=p_{F_{\tau}}}\right]^{-1}
\end{equation}
in nonrelativistic models.
The isoscalar nucleon effective mass $m_{s}^{\ast}$ is the nucleon effective mass in SNM, while the isovector nucleon effective mass $m_{v}^{\ast}$ represents the effective mass of proton (neutron) in pure neutron (proton) matter, and their expressions are provided in Appendix~\ref{sec:Apd_1}.
We introduce a subscript ``0" to denote their values at $\rho_0$, e.g., $m_{s,0}^{\ast}$ and $m_{v,0}^{\ast}$.
Another commonly used quantity is the isospin splitting of nucleon effective mass at the corresponding Fermi momentum, defined as $\msplast(\rho,\delta) \equiv \left[m_{\mathrm{n}}^{\ast}(\rho,\delta)-m_{\mathrm{p}}^{\ast}(\rho,\delta) \right]/m$.
The $\msplast(\rho,\delta)$ can be expanded as a power series in $\delta$, i.e.,
\begin{equation}
\label{eq:spl_coes}
\msplast(\rho,\delta) =\sum_{n=1}^{\infty}
\Delta m_{2 n-1}^{\ast}(\rho) \delta^{2 n-1},
\end{equation}
where the first coefficient $\Delta m_{1}^{\ast}(\rho)$ is usually referred to as the linear isospin splitting coefficient, whose expression is presented in Appendix~\ref{sec:Apd_1}.
Since the effective mass isospin splitting is also momentum-dependent in general, we introduce the notation $\mspl$ for the splitting at arbitrary momentum.

\end{widetext}

\section{Fitting strategy and new interactions}
\label{sec:fitting}
The force-range parameters $\Lambda_i$ in the MD term represent the ranges of the nuclear interaction.
We set $\Lambda_1 = 140$~MeV/$c$, $\Lambda_2 = 600$~MeV/$c$, approximately corresponding to the masses of $\pi$ meson and $\sigma$ (or $\rho$, $\omega$) meson, respectively, to mimic the effective long-range and intermediate-range components.
% Regarding the parameter $\Lambda_3$, which contributes to the short-range force, we will leave its value undetermined for now.
% For the short-range one $\Lambda_3$, we will leave its value undetermined for now.
% This is because we have already introduced the zero-range DD terms, which could also contribute to the short-range force.
The value of $\Lambda_3$, which corresponds to the short-range force, can be determined from the high-energy behavior of the single-nucleon potential, as will be demonstrated later.
After fixing the values of $\Lambda_1$, $\Lambda_2$ and $\Lambda_3$, there are still $14$ parameters in the MDI3Y model, which require $14$ independent quantities to completely construct an interaction.
In Table~\ref{tab:eMDI}, we list the model parameters and the adjustable quantities for the MDI3Y model.

The quantities $\rho_0$, $E_0(\rho_0)$, $K_0$, $J_0$ and $E_{{\mathrm{sym}}}(\rho_0)$, $L$, $K_{{\mathrm{sym}}}$, $J_{{\mathrm{sym}}}$ are related to the bulk properties of the isospin asymmetric nuclear matter.
It is worth noting again that including additional DD terms in Eq.~(\ref{eq:V_DD}) may provide more flexible high-density behavior, which could be useful for extracting the nuclear EOS at suprasaturation densities with new available experimental probes in the future.
The $C_1$, $C_2$, $C_3$ [Eq.~(\ref{eq:Ci})] determine the momentum dependence of $U_0$, which can be obtained by fitting the nucleon optical potential~\cite{Hama:1990vr,Cooper:1993nx}, while $D_1$, $D_2$, $D_3$ [Eq.~(\ref{eq:Di})] determine the momentum dependence of $U_{\mathrm{sym}}$.
On the other hand, the behavior of $U_{\mathrm{sym}}$ can also be characterized by the quantities $\Delta m_{1}^{\ast}(\rho_0)$, $m_{v,0}^{\ast}$, and $\Delta U_{\mathrm{sym}}(\rho)$, with
\begin{equation}
\label{eq:DeltaUsym}
\begin{aligned}
\Delta U_{\mathrm{sym}}(\rho) \equiv & U_{\mathrm{sym}}(\rho,p=0) - U_{\mathrm{sym}}(\rho,p=\infty) \\
=& D_1 \frac{ p_{F}^{3}}{p_{F_0}^{3}} \frac{\Lambda_{1}^2 }{p_{F}^{2}+\Lambda_{1}^{2}}
+ D_2 \frac{ p_{F}^{3}}{p_{F_0}^{3}} \frac{\Lambda_{2}^2 }{p_{F}^{2}+\Lambda_{2}^{2}} \\
&+ D_3 \frac{ p_{F}^{3}}{p_{F_0}^{3}} \frac{\Lambda_{3}^2 }{p_{F}^{2}+\Lambda_{3}^{2}}.
\end{aligned}
\end{equation}
The values of $D_1$, $D_2$, $D_3$ can be uniquely determined by $\Delta m_{1}^{\ast}(\rho_0)$, $m_{v,0}^{\ast}$, and $\Delta U_{\mathrm{sym}}(\rho_0)$ from Eqs.~(\ref{eq:Dm1}), (\ref{eq:mv}), and (\ref{eq:DeltaUsym}).
Due to the lack of experimental data, the behavior of $U_{\mathrm{sym}}$ still exhibits significant uncertainties.
To construct different types of $U_{\mathrm{sym}}$, we can either fit the results from theoretical calculations or adjust the values of $\Delta m_{1}^{\ast}(\rho_0)$, $m_{v,0}^{\ast}$, and $\Delta U_{\mathrm{sym}}(\rho_0)$, and we will discuss these two approaches in detail later.
% The $\Delta m_{1}^{\ast}(\rho_0)$, $m_{v,0}^{\ast}$, and $\Delta U_{\mathrm{sym}}(\rho_0)$ uniquely determine the values of $D_1$, $D_2$, and $D_3$.
%%%%%%%%%%%%%%%%% Table %%%%%%%%%%%%%%%%%%%%%%%%%%%%%%%%%%%%%%%%%
\begin{table}
% \small
\caption{
\label{tab:eMDI}
The 14 model parameters and the adjustable quantities for the MDI3Y model.
}
% \begin{ruledtabular}
\begin{tabular}{|c|c|}
\hline
 Parameters & Quantities  \\ \hline
  $A_l$, $A_u$, $t_{3}^{[1]}$, $x_{3}^{[1]}$, $t_3^{[3]}$, $x_{3}^{[3]}$,
& $\rho_0$, $E_0(\rho_0)$, $K_0$, $J_0$, $E_{{\mathrm{sym}}}(\rho_0)$,
$L$, \\
$t_3^{[5]}$, $x_{3}^{[5]}$, $C_{l,1}$, $C_{u,1}$,
& $K_{{\mathrm{sym}}}$, $J_{{\mathrm{sym}}}$, $C_1$, $C_2$, $C_3$,
[$D_1$, $D_2$, \\
$C_{l,2}$, $C_{u,2}$, $C_{l,3}$, $C_{u,3}$
&  $D_3$ or  $\Delta m_{1}^{\ast}(\rho_0)$, $m_{v,0}^{\ast}$,
       $\Delta U_{\mathrm{sym}}(\rho_0)$]
      \\ \hline
\end{tabular}
% \end{ruledtabular}
\end{table}
%%%%%%%%%%%%%%%%%%%%%%%%%%%%%%%%%%%%%%%%%%%%%%%%%%%%%%%%%%

In the construction of the MDI3Y interactions, we take the values of $\rho_0$, $E_0(\rho_0)$, and $K_0$ to be $0.16$~fm$^{-3}$, $-16$~MeV, and $230$~MeV, respectively.
Next, with different values of $\Lambda_3$, we use GEKKO optimization suite~\cite{beal2018gekko} to minimize the weighted squared difference $\chi^{2}$ between $U_0$ in Eq.~(\ref{eq:U0_C}) and the nucleon optical potential data $U_{\mathrm{opt}}$ up to $1$~GeV obtained by Hama \textit{et al.}~\cite{Hama:1990vr,Cooper:1993nx}:
\begin{equation}
\chi^{2} = \sum_{i=1}^{N_{d}} \left( \frac{U_{0,i}-U_{\mathrm{opt},i}}{\sigma_{i}} \right)^{2},
\end{equation}
with constraint of the Hugenholtz-Van Hove theorem~\cite{Hugenholtz:1958zz,SATPATHY199985}, i.e.,
\begin{equation}
E_{0}\left(\rho_0\right) = \frac{p_{F_0}^{2}}{2m} +U_{0} \left(\rho_0,p_{F_0}\right).
\end{equation}
The $N_d$ is the number of the experimental data points.
Since there are actually no practical errors $\sigma_i$ here, we assign equal weights to each data point.

%%%%%%%%%%%%%%%%% Figure %%%%%%%%%%%%%%%%%%%
\begin{figure}[ht]
    \centering
    \includegraphics[width=\linewidth]{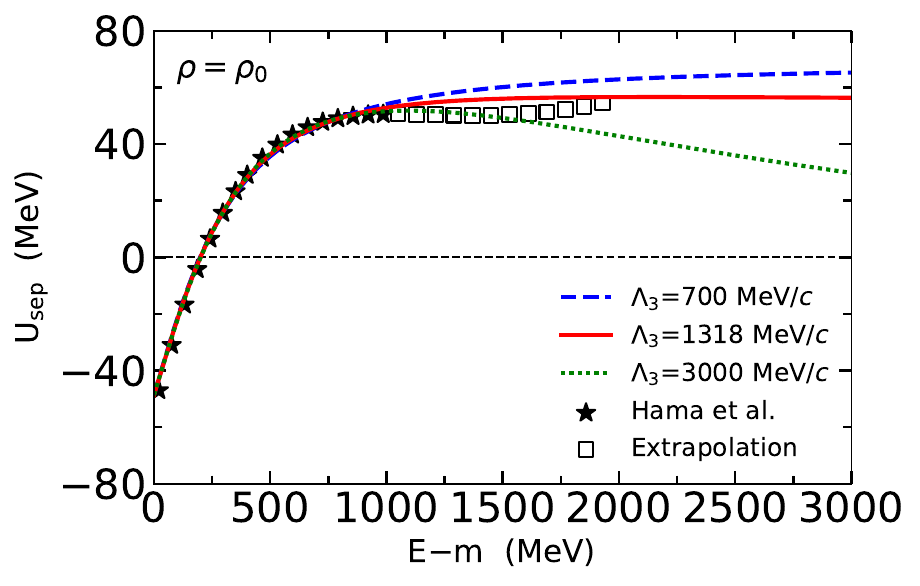}
    \caption{The energy dependence of single-nucleon potential in cold SNM with $\Lambda_3=700$, $1318$, and $3000$~MeV/$c$, respectively, in the MDI3Y model.
    Also shown are the nucleon optical potential (Schr\"{o}dinger equivalent potential, $\mathrm{U}_{\mathrm{sep}}$) in SNM at $\rho_0$ obtained by Hama \textit{et al.}~\cite{Hama:1990vr,Cooper:1993nx}.}
    \label{fig:U0_L3}
\end{figure}
%%%%%%%%%%%%%%%%%%%%%%%%%%%%%%%%%%%%%%%%%%%%

%%%%%%%%%%%%%%%%% Figure %%%%%%%%%%%%%%%%%%%
\begin{figure}[ht]
    \centering
    \includegraphics[width=\linewidth]{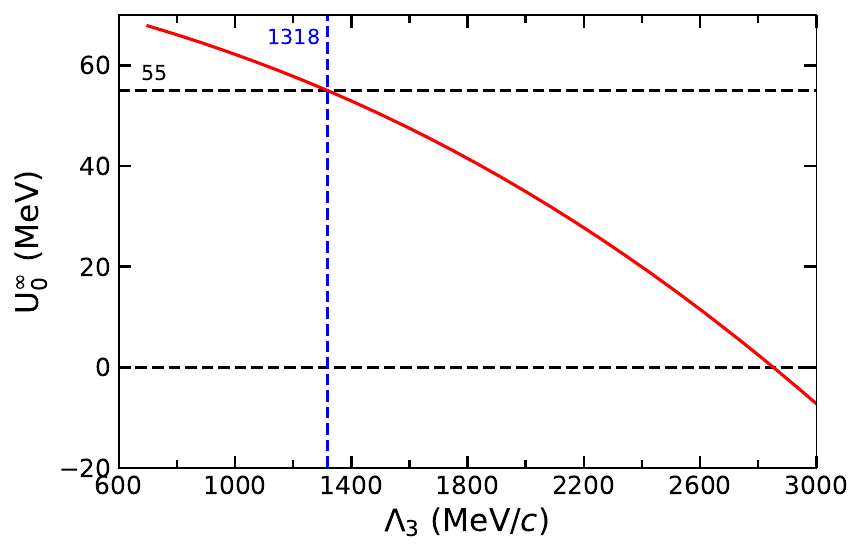}
    \caption{The predicted values of $U_{0}^{\infty}$ for different $\Lambda_3$ in the MDI3Y model.
    Note $U_{0}^{\infty}=55$~MeV with $\Lambda_3=1318$~MeV/$c$.}
    \label{fig:U0inf_L3}
\end{figure}
%%%%%%%%%%%%%%%%%%%%%%%%%%%%%%%%%%%%%%%%%%%%

\begin{figure}[ht]
    \centering
    \includegraphics[width=\linewidth]{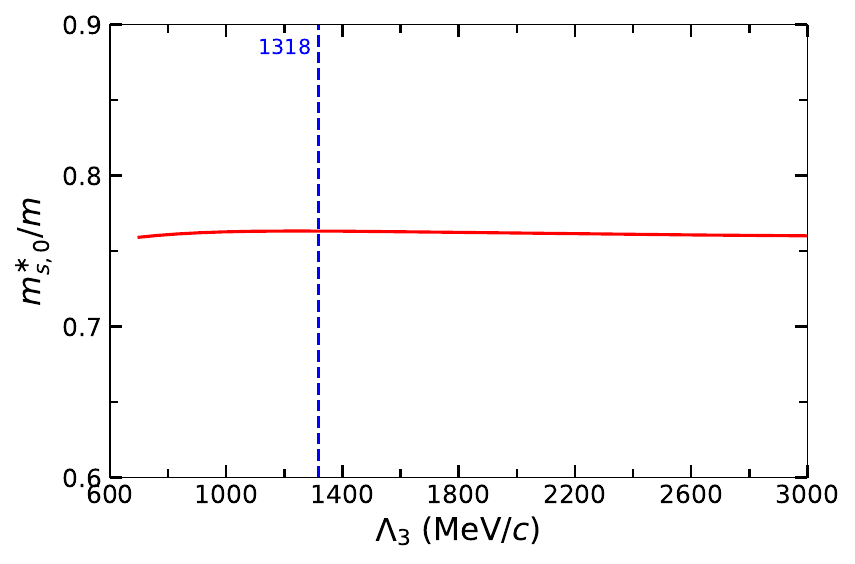}
    \caption{The predicted values of $m_{s,0}^{\ast}$ for different $\Lambda_3$ with the MDI3Y interaction. The dashed line indicates $\Lambda_3=1318$~MeV/$c$ with $m_{s,0}^{\ast}/m=0.763$.}
    \label{fig:ms_L3}
\end{figure}

In Fig.~\ref{fig:U0_L3}, we present the energy dependence of $U_0$ calculated from Eq.~(\ref{eq:U0_C}) using the optimized $C_i$ for different values of $\Lambda_3$.
% which contributes to the effective short-range force ($\Lambda_3>\Lambda_2$).
For illustration, we select $\Lambda_3=700$, $1318$, and $3000$~MeV/$c$ as representative cases.
It can be seen from Fig.~\ref{fig:U0_L3} that the Hama's data can be successfully reproduced with different choices of $\Lambda_3$, while their predictions above $1$~GeV could differ significantly.
From Fig.~\ref{fig:U0_L3}, it can be observed that when $\Lambda_3=1318$~MeV/$c$, the predicted $U_0$ provides a nice description of the extrapolation of Hama's data.
In Fig.~\ref{fig:U0inf_L3}, we show the values of $U_0$ at infinitely large nucleon momentum at $\rho_0$, denoted as $U_{0}^{\infty}$, for different $\Lambda_3$, with $U_{0}^{\infty}=55$~MeV when $\Lambda_3=1318$~MeV/$c$.
In Fig.~\ref{fig:ms_L3}, we present the values of the isoscalar effective mass at $\rho_0$, denoted as $m_{s,0}^{\ast}$, for different $\Lambda_3$.
The value of $m_{s,0}^{\ast}$ is determined by the behavior of $U_0$ around the Fermi momentum, where the predictions for different $\Lambda_3$ are nearly identical, resulting in almost consistent values of $m_{s,0}^{\ast}$.
For $\Lambda_3=1318$~MeV/$c$, $m_{s,0}^{\ast}$ is found to be $0.763m$.
Based on the above discussion, we set $\Lambda_3=1318$~MeV/$c$ throughout this paper to fit the extrapolation of the Hama's data at high energies, and in this case, we obtain that $C_1=106.476$~MeV, $C_2=-249.137$~MeV, and $C_3=67.8626$~MeV.
It should be pointed out that, by changing $\Lambda_3$, we could adjust the value of $U_{0}^{\infty}$, and thus we can investigate the high-energy behaviors of single-nucleon potential, which remains largely uncertain.
So far, the last undetermined quantity related to SNM is the skewness parameter $J_0$, which characterizes the stiffness of $E_0(\rho)$ at suprasaturation densities.
In this work, we set the value of $J_0$ to $-460$~MeV, which is the maximum value that satisfies the constraints on the SNM pressure from flow data in HICs~\cite{Danielewicz:2002pu}.

%%%%%%%%%%%%%%%%% Figure %%%%%%%%%%%%%%%%%%%
%%%%%%%%%%%%%%%%%%%%%%%%%%%%%%%%%%%%%%%%%%%%

In the next step, we discuss the construction of $U_{\mathrm{sym}}$.
Considering its significant uncertainties, we construct four different momentum dependence for $U_{\mathrm{sym}}$, including both the monotonic and non-monotonic types.
The different $U_{\mathrm{sym}}$ are constructed in two ways:
(i) directly fitting the theoretical results up to $800$ MeV/$c$ from relativistic Dirac-Brueckner-Hartree-Fock (DBHF) theory~\cite{vanDalen:2005sk};
(ii) adjusting the values for $\Delta m_{1}^{\ast}(\rho_0)$, $m_{v,0}^{\ast}$, and $\Delta U_{\mathrm{sym}}(\rho_0)$.
For the first approach, the fitting process is the same as that used for determining $U_0$:
we minimize the difference between $U_{\mathrm{sym}}$ in Eq.~(\ref{eq:Usym_D}) and the results of DBHF calculation, and we obtain that $D_1=258.339$~MeV, $D_2=-232.462$~MeV, and $D_3=344.078$~MeV.
The obtained $U_{\mathrm{sym}}$ initially increases and then decreases with increasing momentum, as predicted by the DBHF calculation.
We denote this $U_{\mathrm{sym}}$ type as D017 since the $\Delta m_{1}^{\ast}(\rho_0)$ is obtained as $0.171$, which nicely agrees with result $\msplast(\rho_0,\delta)=0.187\delta$ from the relativistic BHF theory in the full Dirac space~\cite{Wang:2023owh}.
For the second approach, we construct three types of $U_{\mathrm{sym}}$, including two cases for $\Delta m_{1}^{\ast}(\rho_0)=0.2$:
a monotonically decreasing type denoted as D02, and a type that initially decreases and then increases denoted as D02r;
as well as one case for $\Delta m_{1}^{\ast}(\rho_0)=-0.2$ that initially increases and then decreases, denoted as Dm02.
Typically, $\Delta m_{1}^{\ast}(\rho_0)$ is related to the slope of $U_{\mathrm{sym}}$ at $p_{F_0}$, while $\Delta U_{\mathrm{sym}}(\rho_0)$ determines its asymptotic behavior at high momenta.
And $m_{v,0}^{\ast}$ can be adjusted accordingly to modify the structure of $U_{\mathrm{sym}}$.
The values of $\Delta m_{1}^{\ast}(\rho_0)$, $m_{v,0}^{\ast}$, and $\Delta U_{\mathrm{sym}}(\rho_0)$ for D017, D02, D02r, and Dm02 are listed in Table~\ref{tab:Macros}, and the corresponding momentum dependence of $U_{\mathrm{sym}}$ can be found in Fig.~\ref{fig:Usym}.
The value of $D_0$ can be determined through:
\begin{equation}
\label{eq:Esym_decom}
E_{\mathrm{sym}}(\rho_0)=\frac{1}{3} \frac{p_{F_0}^{2}}{2m^{\ast}_{s,0}}+\frac{1}{2} U_{\mathrm{sym}}\left(\rho_0,p_{F_0}\right),
\end{equation}
once $E_{{\mathrm{sym}}}(\rho_0)$ is given~\cite{Brueckner:1964zz,Dabrowski:1972mbb,Dabrowski:1973zz,Xu:2010fh,Xu:2010kf,Chen:2011ag}.

The remaining four undetermined quantities $E_{{\mathrm{sym}}}(\rho_0)$, $L$, $K_{{\mathrm{sym}}}$, and $J_{{\mathrm{sym}}}$ describe the magnitude and density dependence of $E_{{\mathrm{sym}}}(\rho)$.
We first consider the case of $L=35$~MeV, and then we set $E_{{\mathrm{sym}}}(\rho_0)=32$~MeV, $K_{{\mathrm{sym}}}=-275$~MeV, and $J_{{\mathrm{sym}}}=720$~MeV to satisfy the following constrains:
(1) the EOS of PNM obtained from the various microscopic calculations (see Ref.~\cite{Zhang:2022bni} and the references therein);
(2) the largest mass of neutron stars observed to date from PSR J0740+6620~\cite{NANOGrav:2019jur,Fonseca:2021wxt};
(3) the upper limit $\Lambda_{1.4} \leqslant 580$ of the dimensionless tidal deformability of a canonical $1.4$~$M_{\odot}$ neutron star from the gravitational wave event GW170817~\cite{LIGOScientific:2018cki};
(4) the mass-radius (M-R) relation for PSR J0030+0451~\cite{Miller:2019cac,Riley:2019yda} with a mass of $1.4$~$M_{\odot}$, PSR J0437-4715~\cite{Choudhury:2024xbk} with a mass of $1.4$~$M_{\odot}$, PSR J0740+6620~\cite{Miller:2021qha,Riley:2021pdl} with a mass of $2.0$~$M_{\odot}$, and the central compact object (CCO) within HESS J1731-347~\cite{Doroshenko:2022nwp} with an unusually low mass of $0.77$~$M_{\odot}$ and small radius of $10.4$~km.

By combining the $E_{{\mathrm{sym}}}(\rho_0)$, $L$, $K_{{\mathrm{sym}}}$, and $J_{{\mathrm{sym}}}$ with four types of $U_{\mathrm{sym}}$ and the foregoing isoscalar quantities, we have constructed four different interactions, denoted as MDI3YL35D017, MDI3YL35D02, MDI3YL35D02r, and MDI3YL35Dm02, respectively.
To further investigate the effects of $E_{{\mathrm{sym}}}$, we provide two additional quantities sets:
$L=55$~MeV, $E_{{\mathrm{sym}}}(\rho_0)=33$~MeV, $K_{{\mathrm{sym}}}=-100$~MeV, $J_{{\mathrm{sym}}}=1200$~MeV;
and $L=75$~MeV, $E_{{\mathrm{sym}}}(\rho_0)=34$~MeV, $K_{{\mathrm{sym}}}=-80$~MeV, $J_{{\mathrm{sym}}}=1000$~MeV.
For these two quantity sets of the symmetry energy, we select the D02 type of $U_{\mathrm{sym}}$ to construct two additional interactions, labeled as MDI3YL55D02 and MDI3YL75D02.
Based on the MDI3Y model, we have thus constructed six interactions in this work.
It is worth emphasizing again that these six interactions have the same description of the SNM properties.
The four interactions MDI3YL35D017, MDI3YL35D02, MDI3YL35D02r, and MDI3YL35Dm02 exhibit similar density behavior of the symmetry energy, while the MDI3YL35D02, MDI3YL55D02, and MDI3YL75D02 share the same momentum dependence of $U_{\mathrm{sym}}$.
For brevity, we adopt the convention MDI3YX to label different interactions in subsequent tables and figures, with X indicating both
the value of $L$ and the symmetry potential type.
In Table~\ref{tab:Paras_vals}, we list the 14 model parameters for these MDI3Y interactions.

We would like to point out that the isoscalar quantities have significant impact on the nuclear EOS as well as neutron star properties. Here we fix the isoscalar part to focus on the effects of symmetry potential, which are designed for future HICs simulations to extract some important isovector quantities such as the isospin splitting of nucleon effective mass. However, when constraining the nuclear EOS using neutron star and/or HICs data through Bayesian inference, both the isoscalar and isovector parameters should be treated as free variables~\cite{Davis:2024nda}.

%%%%%%%%%%%%%%%%% Table %%%%%%%%%%%%%%%%%%%%%%%%%%%%%%%%%%%%%%%%%%%%%%%%%%%%%%%%%%%%%%%%%%%%%
\begin{table*}
\caption{\label{tab:Paras_vals}
The 14 model parameters of the MDI3YX interactions.
$x_{3}^{[n]}$ ($n=1,3,5$) are dimensionless.
For all interactions: $\Lambda_1=140$~MeV/$c$, $\Lambda_2=600$~MeV/$c$, $\Lambda_3=1318$~MeV/$c$, $C_{1}=C_{l,1}+C_{u,1}=106.476$~MeV, $C_{2}=C_{l,2}+C_{u,2}=-249.137$~MeV, $C_{3}=C_{l,3}+C_{u,3}=67.8626$~MeV.
}
\begin{ruledtabular}
\begin{tabular}{ccccccc}
X=& L35D017& L35D02& L35D02r& L35Dm02& L55D02& L75D02  \\ \hline
$A_l~(\rm{MeV})$
& -340.957 & -153.841 & -48.9242 & -115.571 & -139.574  & -220.641
\\
$A_u~(\rm{MeV})$
& 107.377 & -79.7389 & -184.655 & -118.008 & -94.0055  & -12.9389
\\
$C_{l,1}~(\rm{MeV})$
& 182.407 & 85.1073 & 60.0178 & 57.0134 & 85.1073   & 85.1073
\\
$C_{u,1}~(\rm{MeV})$
& -75.9315 & 21.3685 & 46.4581 & 49.4624 & 21.3685   & 21.3685
\\
$C_{l,2}~(\rm{MeV})$
& -240.799 & -103.423 & -53.0583 & -190.268 & -103.423  & -103.423
\\
$C_{u,2}~(\rm{MeV})$
& -8.33773 & -145.713 & -196.078 & -58.8684 & -145.713  & -145.713
\\
$C_{l,3}~(\rm{MeV})$
& 205.970 & 36.7689 & -50.7926 & 131.961 & 36.7689   & 36.7689
\\
$C_{u,3}~(\rm{MeV})$
& -138.107 & 31.0936 & 118.655 & -64.0988 & 31.0936   & 31.0936
\\
$t_{3}^{[1]}~(\mathrm{MeV}\, \mathrm{fm}^{4})$
& 16321.3   & 16321.3  & 16321.3   & 16321.3   & 16321.3   & 16321.3
\\
$t_{3}^{[3]}~(\mathrm{MeV}\, \mathrm{fm}^{6})$
& -9094.38 & -9094.38  & -9094.38  & -9094.38  & -9094.38  & -9094.38
\\
$t_{3}^{[5]}~(\mathrm{MeV}\, \mathrm{fm}^{8})$
& 5024.65  & 5024.65   & 5024.65   & 5024.65   & 5024.65   & 5024.65
\\
$x_{3}^{[1]}$
& -1.93719 & -1.33714 & -1.22204 & -0.578062 & -1.11707  & -2.08202
\\
$x_{3}^{[3]}$
& -7.99556 & -5.87032 & -5.43377 & -5.02137 & -5.64697  & -7.57124
\\
$x_{3}^{[5]}$
& -11.6092 & -8.49551 & -7.74781 & -7.84116 & -11.4709  & -13.4967
\end{tabular}
\end{ruledtabular}
\end{table*}
%%%%%%%%%%%%%%%%%%%%%%%%%%%%%%%%%%%%%%%%%%%%%%%%%%%%%%%%%%%%%%%%%%%%%%%%%%%%%%%%%%%%%%

\section{The properties of cold nuclear matter}
\label{sec:properties}
Table~\ref{tab:Macros} summarizes the characteristic quantities of nuclear matter obtained with the six MDI3Y interactions constructed above.
It is seen that all the six interactions have the same isoscalar properties as imposed in prior. For the four interactions with $L=35$~MeV, they predict different values for the higher-order parameters $I_{{\rm{sym}}}$,$H_{{\rm{sym}}}$, $E_{{\rm{sym}},4}(\rho_0)$ and $U_{\mathrm{sym}}^{\infty}$, due to the different momentum dependence of $U_{\rm sym}$ by using different $\Delta m_1^{\ast}(\rho_0)$, $m^\ast_{v,0}/m$ and $\Delta U_{\mathrm{sym}}(\rho_0)$.

%%%%%%%%%%%%%%%%% Table %%%%%%%%%%%%%%%%%%%%%%%%%%%%%%%%%%%%%%%%%%%%%%%%%%%%%%%%%%%%%%%%%%%%%
\begin{table*}
\caption{\label{tab:Macros}
Characteristic quantities of nuclear matter with the MDI3YX interactions.
}
\begin{ruledtabular}
\begin{tabular}{ccccccc}
X=& L35D017& L35D02 & L35D02r & L35Dm02& L55D02& L75D02  \\ \hline
$\rho_0\,(\rm{fm}^{-3})$ &
0.160   & 0.160  & 0.160   & 0.160   & 0.160   & 0.160
\\
$E_0(\rho_0)\,(\rm{MeV})$ &
-16.0   & -16.0  & -16.0  & -16.0   & -16.0   & -16.0
\\
$K_0\,(\rm{MeV})$ &
230.0   & 230.0  & 230.0   & 230.0   & 230.0   & 230.0
\\
$J_0\,(\rm{MeV})$ &
-460.0  & -460.0 & -460.0  & -460.0  & -460.0  & -460.0
\\
$I_0\,(\rm{MeV})$ &
2228.7  & 2228.7  & 2228.7  & 2228.7  & 2228.7  & 2228.7
\\
$H_0\,(\rm{MeV})$ &
-12841  & -12841  & -12841  & -12841  & -12841  & -12841
\\
$m^\ast_{s,0}/m$ &
 0.763  & 0.763  & 0.763   & 0.763   & 0.763   & 0.763
\\
$U_{0}^{\infty}\,(\mathrm{MeV})$ &
55.0 &55.0 &55.0 &55.0 &55.0 &55.0
\\
$E_{{\rm{sym}}}(\rho_0)\,(\rm{MeV})$ &
32.0   & 32.0  & 32.0   & 32.0   & 33.0  & 34.0
\\
$L\,(\rm{MeV})$ &
35.0  & 35.0   & 35.0   & 35.0   & 55.0  & 75.0
\\
$K_{{\rm{sym}}}\,(\rm{MeV})$ &
-275  & -275 & -275  & -275  & -100  & -80.0
\\
$J_{{\rm{sym}}}\,(\rm{MeV})$ &
720.0   & 720.0  & 720.0   & 720.0   & 1200    & 1000
\\
$I_{{\rm{sym}}}\,(\rm{MeV})$ &
1454.8 & 237.43 & -87.192 & 113.64 & -658.56 & 1365.4
\\
$H_{{\rm{sym}}}\,(\rm{MeV})$ &
-11342 & -3885.8 & -1851.6 & -3178.5 & 1778.1   & -15437
\\
$E_{{\rm{sym}},4}(\rho_0)\,(\rm{MeV})$ &
0.1736 & 1.030 & 1.265 & 0.5356 & 1.030 & 1.030
\\
$\Delta m_1^{\ast}(\rho_0)$ &
0.17 & 0.20 & 0.20 & -0.20  & 0.20 & 0.20
\\
$m^\ast_{v,0}/m$ &
0.606 & 0.575 & 0.565 & 0.790 & 0.575   & 0.575
\\
$\Delta U_{\mathrm{sym}}(\rho_0)\,(\mathrm{MeV})$ &
193 & 55.0 & -40.0 & 80.0 & 55.0 & 55.0
\\
$U_{\mathrm{sym}}^{\infty}\,(\mathrm{MeV})$ &
-164.7 & -17.08 & 80.34 & -54.08 & -15.08 & -13.08
\end{tabular}
\end{ruledtabular}
\end{table*}
%%%%%%%%%%%%%%%%%%%%%%%%%%%%%%%%%%%%%%%%%%%%%%%%%%%%%%%%%%%%%%%%%%%%%%%%%%%%%%%%%%%%%%

\subsection{Single-nucleon potential, symmetry potential and the nucleon effective mass}

The single-nucleon potential in SNM at $\rho_0$, $U_{0}(\rho_0,p)$, predicted by the MDI3Y interactions, is plotted in Fig.~\ref{fig:U0_L3} (in red) as a function of the nucleon kinetic energy $E-m=\sqrt{p^2+m^2}+U_{0}(\rho,p)-m$.
Note the six MDI3Y interactions have the same $\Lambda_3=1318$~MeV/$c$.
Fig.~\ref{fig:U0_L3} additionally includes the real part of the nucleon optical potential (Schr\"{o}dinger equivalent potential) obtained by Hama \textit{et al.}~\cite{Hama:1990vr,Cooper:1993nx} based on Dirac phenomenology of nucleon-nucleus scattering data.
It is seen that the predicted $U_{0}(\rho_0,p)$ agrees well with the empirical nucleon optical potential obtained by Hama \textit{et al.} for nucleon kinetic energy up to $1$~GeV and also matches its extrapolation at higher energies.
The value of $U_{0}(\rho_0,p)$ at infinitely large nucleon momentum, $U_{0}^{\infty}$, is obtained as $55$~MeV.

%%%%%%%%%%%%%%%%% Figure %%%%%%%%%%%%%%%%%%%
\begin{figure}[ht]
    \centering
    \includegraphics[width=\linewidth]{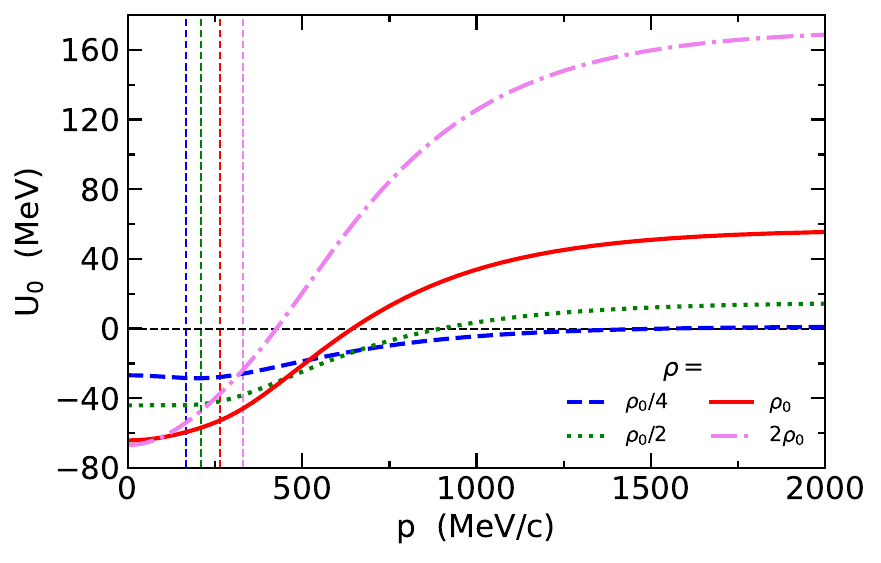}
    \caption{Momentum dependence of the single-nucleon potential in SNM $U_{0}(\rho,p)$, at $\rho=\rho_0/4$, $\rho_0/2$, $\rho_0$, and $2\rho_0$, respectively.
    The vertical lines indicate the corresponding Fermi momenta.}
    \label{fig:U0_p}
\end{figure}
%%%%%%%%%%%%%%%%%%%%%%%%%%%%%%%%%%%%%%%%%%%%

Figure~\ref{fig:U0_p} displays the momentum dependence of $U_{0}(\rho,p)$ at $\rho=\rho_0/4$, $\rho_0/2$, $\rho_0$, and $2\rho_0$, respectively, with the corresponding Fermi momentum $p_{F}$ indicated in the plot.
The momentum dependence of $U_{0}(\rho,p)$ at $p_{F}$ can be characterized by the isoscalar nucleon effective mass $m_{s}^{\ast}(\rho)$, as shown in Eq.~(\ref{eq:ms_exp}).
It is seen that $U_{0}(\rho_0/4,p)$ is decreasing with nucleon momentum at $p_{F}$, which consequently leads to $m_{s}^{\ast}(\rho_0/4)/m>1$.

It is interesting to mention that $m_{s}^{\ast}$ is also related to the single-particle states in finite nuclei.
From mean-field calculation, it has been found that $m_{s}^{\ast}/m \approx 1$ is necessary to reproduce the level density near the Fermi surface~\cite{BROWN1963598,Barranco:1980km,Shlomo:1992zcu,Mughabghab:1998zz} and to accurately fit the masses of open-shell nuclei~\cite{Tondeur:2000bd,Goriely:2001zz}, while $m_{s}^{\ast}/m$ should be about $0.7$ for deeper states~\cite{Farine:2001oac}.
On the other hand, the value of $m_{s}^{\ast}/m \approx 0.8 \pm 0.1$ is supported by both the analyses of isoscalar giant quadrupole resonances and various many-body calculations (see Ref.~\cite{Li:2018lpy} for a review).
This inconsistent large value of $m_{s}^{\ast}$ near the Fermi surface could be explained by the coupling between particle modes and surface-vibration modes, which leads to modifications of the single-particle energies near the Fermi surface and reducing the energy gap between occupied and unoccupied shells~\cite{Bertsch:1968qxf,Hamamoto:1976quz,Bernard:1980gpb,Ma:1983aqb,Litvinova:2006ds}.
To simultaneously describe the single-particle dynamics near and far from the Fermi surface, a radius-dependent $m_{s}^{\ast}/m$ has been introduced in Ref.~\cite{Ma:1983aqb}, which is about $0.7$ in the interior and has a peak value of about $1.2$-$1.5$ before becomes $1$ at the surface due to the coupling.
This surface-peaked $m_{s}^{\ast}/m$ was later confirmed in self-consistent Skyrme-Hartree-Fock (SHF) calculations by including new terms into the Skyrme force~\cite{Farine:2001oac,Zalewski:2010ni}.

\begin{figure}[ht]
    \centering
    \includegraphics[width=\linewidth]{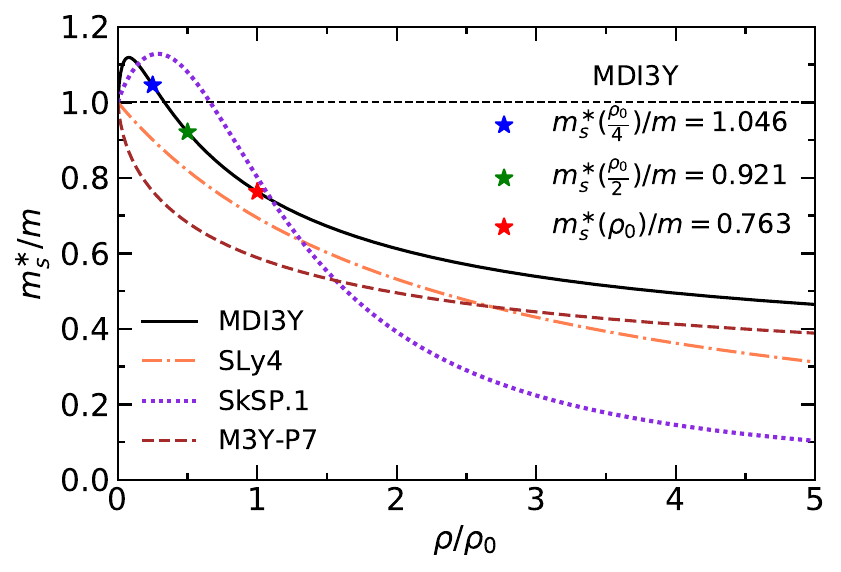}
    \caption{The isoscalar nucleon effective mass $m_{s}^{\ast}(\rho)$ as a function of nucleon density.
    The results for interactions SLy4~\cite{Chabanat:1997un}, SkSP.1~\cite{Farine:2001oac}, and M3Y-P7~\cite{Nakada:2012sq} are also shown for comparison.}
    \label{fig:Ms_rho}
\end{figure}

Shown in Fig.~\ref{fig:Ms_rho} is $m_{s}^{\ast}(\rho)$ as a function of nucleon density for the MDI3Y interactions.
The values of $m_{s}^{\ast}(\rho)$ at $\rho=\rho_0/4$, $\rho_0/2$, and $\rho_0$ are also marked in Fig.~\ref{fig:Ms_rho}.
It is worth noting that these MDI3Y interactions predict the value of $m_{s}^{\ast}(\rho)/m > 1$ for $\rho < \rho_0/3$ with $m_{s}^{\ast}(\rho_0)/m=0.763$.
Also shown in Fig.~\ref{fig:Ms_rho} are the $m_{s}^{\ast}(\rho)$ obtained from interactions SLy4~\cite{Chabanat:1997un}, SkSP.1~\cite{Farine:2001oac}, and M3Y-P7~\cite{Nakada:2012sq}.
By introducing a zero-range term with both momentum and density dependence into the conventional Skyrme interaction, SkSP.1 interaction predicts a surface-peaked $m_{s}^{\ast}(\rho)$ with the value of $m_{s}^{\ast}(\rho_0)/m=0.8$, which may allow for the simultaneous descriptions of both nuclei masses and isoscalar giant quadrupole resonance~\cite{Farine:2001oac}.

%%%%%%%%%%%%%%%%% Figure %%%%%%%%%%%%%%%%%%%
%%%%%%%%%%%%%%%%%%%%%%%%%%%%%%%%%%%%%%%%%%%%

%%%%%%%%%%%%%%%%% Figure %%%%%%%%%%%%%%%%%%%
\begin{figure}[ht]
    \centering
    \includegraphics[width=\linewidth]{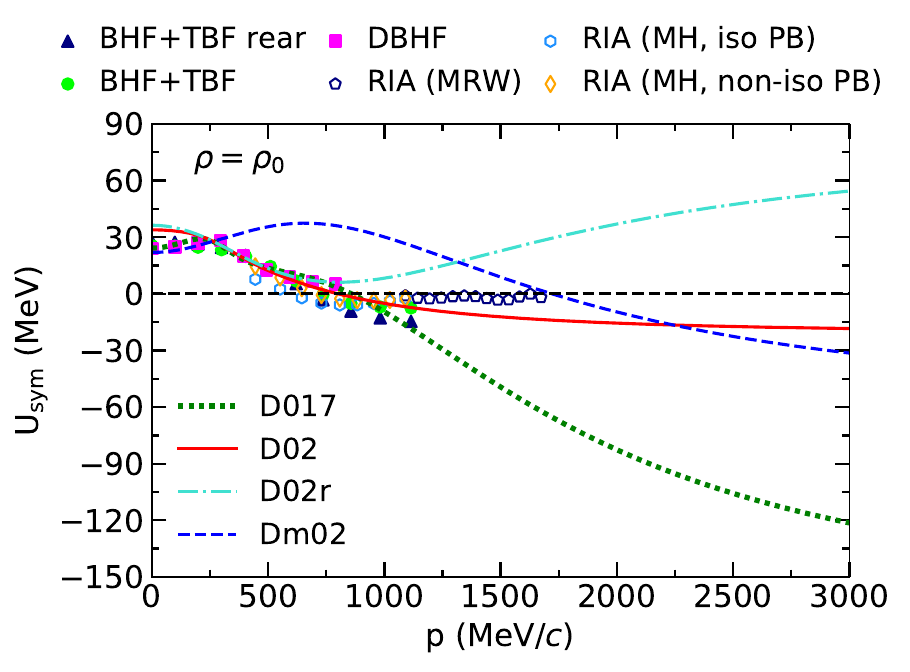}
    \caption{
    Momentum dependence for the four types of $U_{{\mathrm{sym}}}(\rho_0,p)$.
    Here $E_{{\mathrm{sym}}}(\rho_0)=30$~MeV has been assumed.
    The corresponding results from several microscopic calculations are also shown as symbols (see the text for details).}
    \label{fig:Usym}
\end{figure}
%%%%%%%%%%%%%%%%%%%%%%%%%%%%%%%%%%%%%%%%%%%%

Shown in Fig.~\ref{fig:Usym} are the four different momentum dependence of $U_{{\mathrm{sym}}}$ constructed in Sec.~\ref{sec:fitting}.
It is worth emphasizing again that, in the construction of these interactions, the momentum dependence of $U_{{\mathrm{sym}}}$ is solely determined by a set of $\Delta m_{1}^{\ast}(\rho_0)$, $m_{v,0}^{\ast}$, and $\Delta U_{\mathrm{sym}}(\rho_0)$, and is not affected by the properties of $E_{{\mathrm{sym}}}(\rho)$.
Changing the value of $E_{{\mathrm{sym}}}(\rho_0)$ results in a vertical shift of $U_{{\mathrm{sym}}}(\rho_0,p)$, and in Fig.~\ref{fig:Usym}, we have chosen $E_{{\mathrm{sym}}}(\rho_0)=30$~MeV for illustration.
Also shown in Fig.~\ref{fig:Usym} are results from several microscopic calculations, including the non-relativistic BHF theory with and without rearrangement contribution from the three-body force~\cite{Zuo:2006nz}, the relativistic DBHF theory~\cite{vanDalen:2005sk} and the relativistic impulse approximation (RIA)~\cite{Chen:2005hw,Li:2006nd}.
The non-monotonic $U_{{\mathrm{sym}}}$ for D017 are obtained by fitting the results of DBHF calculations.

It should be noted that the behavior of $U_{{\mathrm{sym}}}$ is related to the nucleon effective mass splitting $\mspl$.
Specifically, a decreasing $U_{\mathrm{sym}}$ indicates that in the neutron-rich environments, neutrons have a larger effective mass than protons ($\mspl>0$), whereas an increasing $U_{\mathrm{sym}}$ yields a negative $\mspl$.
The non-monotonic $U_{\mathrm{sym}}$ implies that $\mspl$ can have different signs at different momenta.
Very recently, Sun \textit{et al.}~\cite{Sun:2025ubn} discussed the use of Gogny-type interactions in the QMD models to reproduce the non-monotonic behavior of $U_{\mathrm{sym}}$~\cite{Berger:1991zza,Blaizot:1995zz,Chen:2011ag}.
Actually, transport models predict that neutrons or protons with the same momentum are more likely to be accelerated from the compressed participant region if they experience a stronger repulsive potential and exhibit a smaller effective mass, suggesting that the spectra of emitted particles, such as the neutron over proton spectral ratio, could be a sensitive probe for $\mspl$ and the momentum dependence of $U_{\mathrm{sym}}$~\cite{Rizzo:2005mk,Giordano:2010pv,Zhang:2014sva,Coupland:2014gya,Wang:2023buv,Yang:2025aia}.

%%%%%%%%%%%%%%%%% Figure %%%%%%%%%%%%%%%%%%%
\begin{figure}[ht]
    \centering
    \includegraphics[width=0.95\linewidth]{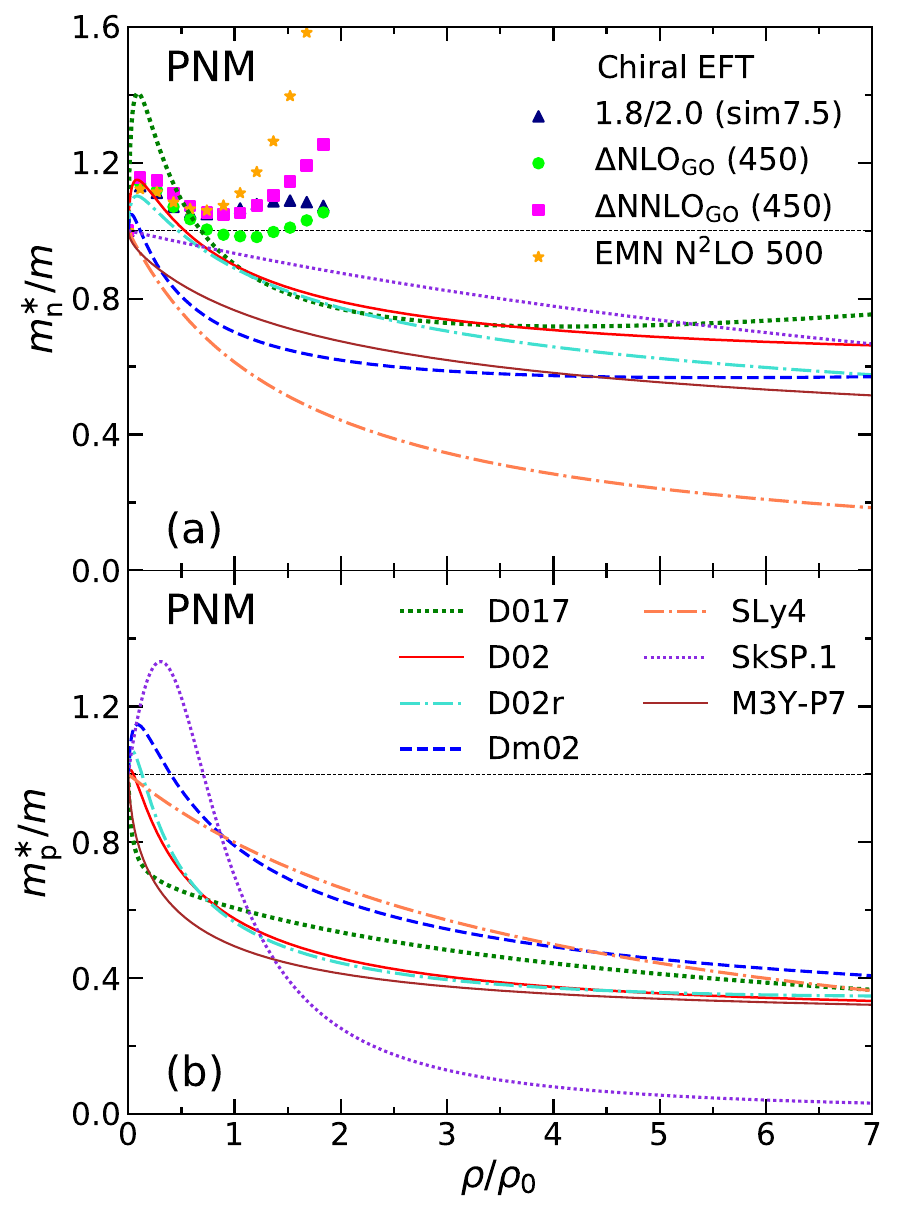}
    \caption{
    Effective masses for neutrons $m_{\mathrm{n}}^{\ast}$ [panel (a)] and protons $m_{\mathrm{p}}^{\ast}$ [panel (b)] in PNM as a function of nucleon density for the four different types of the $U_{{\mathrm{sym}}}$.
    The results for SLy4~\cite{Chabanat:1997un}, SkSP.1~\cite{Farine:2001oac}, and M3Y-P7~\cite{Nakada:2012sq} are included for comparison.
    The $m_{\mathrm{n}}^{\ast}$ in PNM from many-body calculations~\cite{Alp:2025wjn} using various chiral EFT interactions are also shown as solid symbols.
    }
    \label{fig:mtauPNM}
\end{figure}
%%%%%%%%%%%%%%%%%%%%%%%%%%%%%%%%%%%%%%%%%%%%

Figure~\ref{fig:mtauPNM} shows the density dependence of the nucleon effective mass in PNM for the four different types of $U_{{\mathrm{sym}}}$, with the results of neutrons $m_{\mathrm{n}}^{\ast}$ and protons $m_{\mathrm{p}}^{\ast}$ displayed in Fig.~\ref{fig:mtauPNM}(a) and (b), respectively.
Note that the $m_{\mathrm{p}}^{\ast}$ in PNM is equal to the isovector nucleon effective mass $m_{v}^{\ast}$ by definition.
For comparison, Fig.~\ref{fig:mtauPNM} also includes the results for SLy4~\cite{Chabanat:1997un}, SkSP.1~\cite{Farine:2001oac}, and M3Y-P7~\cite{Nakada:2012sq}.
The $m_{\mathrm{n}}^{\ast}$ from the recent many-body calculations~\cite{Alp:2025wjn} using various chiral effective field theory (EFT) interactions (see the reference for details) are also plotted in Fig.~\ref{fig:mtauPNM}(a).
For D017, D02, and D02r with $\Delta m_1^{\ast}(\rho_0)>0$, $m_{\mathrm{n}}^{\ast}$ is always larger than $m_{\mathrm{p}}^{\ast}$.
While for Dm02 with $\Delta m_1^{\ast}(\rho_0)=-0.2$, $m_{\mathrm{n}}^{\ast}$ is smaller than $m_{\mathrm{p}}^{\ast}$ at densities below about $2\rho_0$, but it exceeds $m_{\mathrm{p}}^{\ast}$ at higher densities.

From Fig.~\ref{fig:mtauPNM}(a), it can be seen that $m_{\mathrm{n}}^{\ast}$ exhibits a peak behavior at low densities for D017, D02, D02r, and Dm02, and the peak is most distinct for D017 and least evident for Dm02.
In contrast, $m_{\mathrm{n}}^{\ast}$ is continuously decreasing for SLy4, SkSP.1 and M3Y-P7.
For D017, D02, D02r, Dm02, and M3Y-P7, $m_{\mathrm{n}}^{\ast}$ and $m_{\mathrm{p}}^{\ast}$ will eventually approach to the bare mass in the limit of nucleon density approaching infinity, since they are obtained from finite-range interactions.
It is seen that $m_{\mathrm{n}}^{\ast}$ for D017 and Dm02 begin to increase when $\rho \gtrsim 4\rho_0$.
From Fig.~\ref{fig:mtauPNM}(a), a common peak behavior of $m_{\mathrm{n}}^{\ast}$ is observed below approximately $0.5\rho_0$ for different chiral EFT interactions, although their predictions differ significantly with increasing density.
Note that all the chiral EFT interactions predict an increasing $m_{\mathrm{n}}^{\ast}$ at larger densities, due to the contributions from three-nucleon interactions~\cite{Alp:2025wjn}.
From Fig.~\ref{fig:mtauPNM}(b), it can be observed that $m_{\mathrm{p}}^{\ast}$ exhibits a distinct peak behavior at low densities for Dm02 and SkSP.1.
Compared to SLy4, the peak behavior for SkSP.1 arises from the momentum-dependent DD term, whose contribution also leads to a rapid decrease in $m_{\mathrm{p}}^{\ast}$ with increasing density.

%%%%%%%%%%%%%%%%% Figure %%%%%%%%%%%%%%%%%%%
\begin{figure}[ht]
    \centering
    \includegraphics[width=\linewidth]{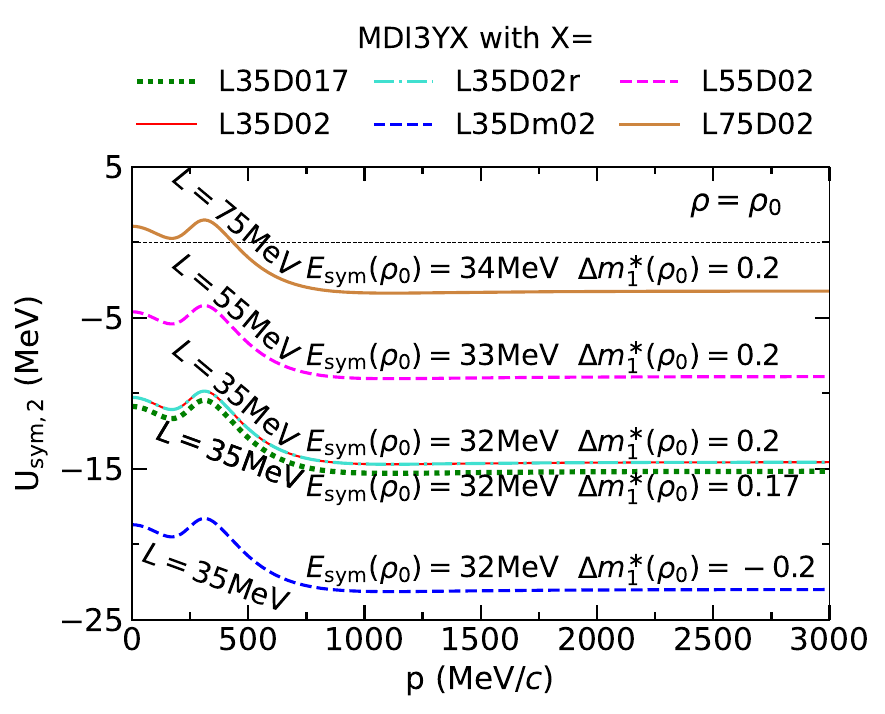}
    \caption{Momentum dependence of the second-order symmetry potential $U_{\mathrm{sym},2}(\rho_0,p)$, obtained with the six MDI3Y interactions.
    Note that the curves of MDI3YL35D02 and MDI3YL35D02r overlap each other.}
    \label{fig:Usym2}
\end{figure}
%%%%%%%%%%%%%%%%%%%%%%%%%%%%%%%%%%%%%%%%%%%%

In Fig.~\ref{fig:Usym2}, we show the second-order symmetry potential $U_{\mathrm{sym},2}$ as a function of nucleon momentum, obtained with the six MDI3Y interactions.
The momentum dependence of $U_{\mathrm{sym},2}$ is determined by the $C_i$ parameters as shown in Eq.~(\ref{eq:Usym2}), which have already been obtained in Sec.~\ref{sec:fitting} by fitting $U_0$ to the empirical nucleon optical potential, and all the $U_{\mathrm{sym},2}$ exhibit the same momentum dependence.
Furthermore, the value of $U_{\mathrm{sym},2}(\rho_0,p_{F_0})$ can be obtained through the single-nucleon potential decomposition of $L$~\cite{Chen:2011ag}:
\begin{equation}
\small
\label{eq:L_HVH}
\begin{aligned}
L= & \frac{2}{3}\frac{p_{F_0}^2}{2m_{s,0}^{\ast}} -\frac{1}{6} \frac{p_{F_0}^3}{(m_{s,0}^{\ast})^2} \left. \frac{\partial m_{s,0}}{\partial p} \right|_{p=p_{F_0}}  + \frac{3}{2} U_{\mathrm{sym},1}(\rho_0,p_{F_0}) \\
& + p_{F_0}  \left. \frac{\partial U_{\mathrm{sym},1}(\rho_0,p) }{\partial p} \right|_{p=p_{F_0}} + 3 U_{\mathrm{sym},2}(\rho_0,p_{F_0}).
\end{aligned}
\end{equation}
Note that Eq.~(\ref{eq:L_HVH}) is shown here at $\rho_0$ but it is actually valid for arbitrary density~$\rho$~\cite{Chen:2011ag}.
Considering that the magnitude of $U_{\mathrm{sym},2}$ is comparable to that of $U_{\mathrm{sym},1}$, the Lane potential~\cite{Lane:1962zz} could be a good approximation for $U_\tau$ only when the isospin asymmetry $\delta$ is small, and the contribution of $U_{\mathrm{sym},2}$ to $L$ should not be neglected, which has been demonstrated in Ref.~\cite{Chen:2011ag}.
Moreover, considering that the $\partial m_{s,0}/ \partial p$ is usually small, the magnitude of $U_{\mathrm{sym},2}$ could be used to evaluate $\Delta m_{1}^{\ast}(\rho_0)$ once the values of $E_{\mathrm{sym}}(\rho_0)$ and $L$ can be relatively well constrained.

\subsection{Bulk properties of cold nuclear matter}
%%%%%%%%%%%%%%%%% Figure %%%%%%%%%%%%%%%%%%%
\begin{figure}[ht]
    \centering
    \includegraphics[width=\linewidth]{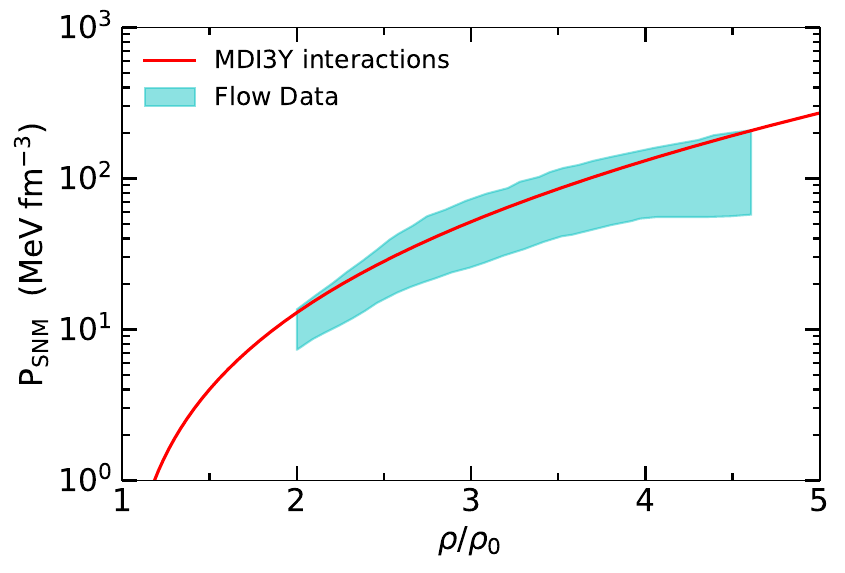}
    \caption{The pressure of SNM $P_{\mathrm{SNM}}(\rho)$ as a function of nucleon density predicted by the MDI3Y interactions.
    The constraints from the flow data in HICs \cite{Danielewicz:2002pu} are also included.}
    \label{fig:Psnm}
\end{figure}
%%%%%%%%%%%%%%%%%%%%%%%%%%%%%%%%%%%%%%%%%%%%

Shown in Fig.~\ref{fig:Psnm} is the density dependence of pressure of SNM $P_{\mathrm{SNM}}(\rho)$ predicted by MDI3Y interactions, and the constraints from flow data in HICs~\cite{Danielewicz:2002pu} are also plotted.
It is seen that all the 6 MDI3Y interactions predict the same $P_{\mathrm{SNM}}(\rho)$, and they all satisfy the constraints from flow data by construction.

%%%%%%%%%%%%%%%%% Figure %%%%%%%%%%%%%%%%%%%
\begin{figure}[ht]
    \centering
    \includegraphics[width=\linewidth]{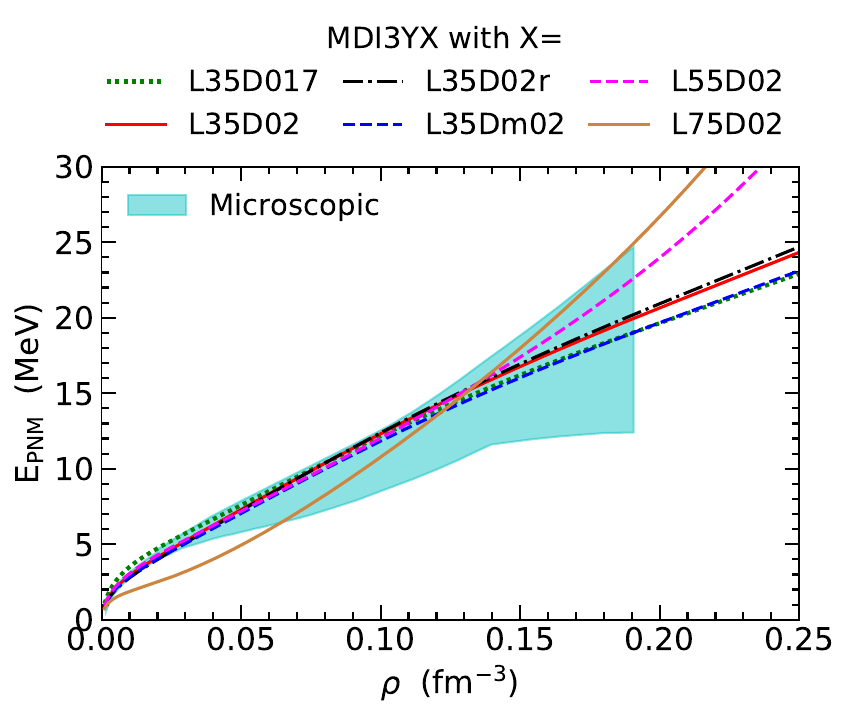}
    \caption{The EOS of PNM $E_{\mathrm{PNM}}(\rho)$ as a function of nucleon density predicted by the MDI3Y interactions.
    Also included are the combined results from various microscopic calculations~\cite{Huth:2020ozf,Zhang:2022bni}.}
    \label{fig:Epnm}
\end{figure}
%%%%%%%%%%%%%%%%%%%%%%%%%%%%%%%%%%%%%%%%%%%%

Shown in Fig.~\ref{fig:Epnm} is the EOS of PNM, i.e., energy per nucleon $E_{\mathrm{PNM}}(\rho)$, as a function of nucleon density predicted by the MDI3Y interactions.
Also included in Fig.~\ref{fig:Epnm} are the combined results from various microscopic calculations~\cite{Huth:2020ozf,Zhang:2022bni}.
It is seen that all interactions, except for MDI3YL75D02, are consistent with the microscopic calculations, whereas MDI3YL75D02 predicts too small $E_{\mathrm{PNM}}$ below about $0.07$~fm$^{-3}$.

%%%%%%%%%%%%%%%%% Figure %%%%%%%%%%%%%%%%%%%
\begin{figure}[ht]
    \centering
    \includegraphics[width=\linewidth]{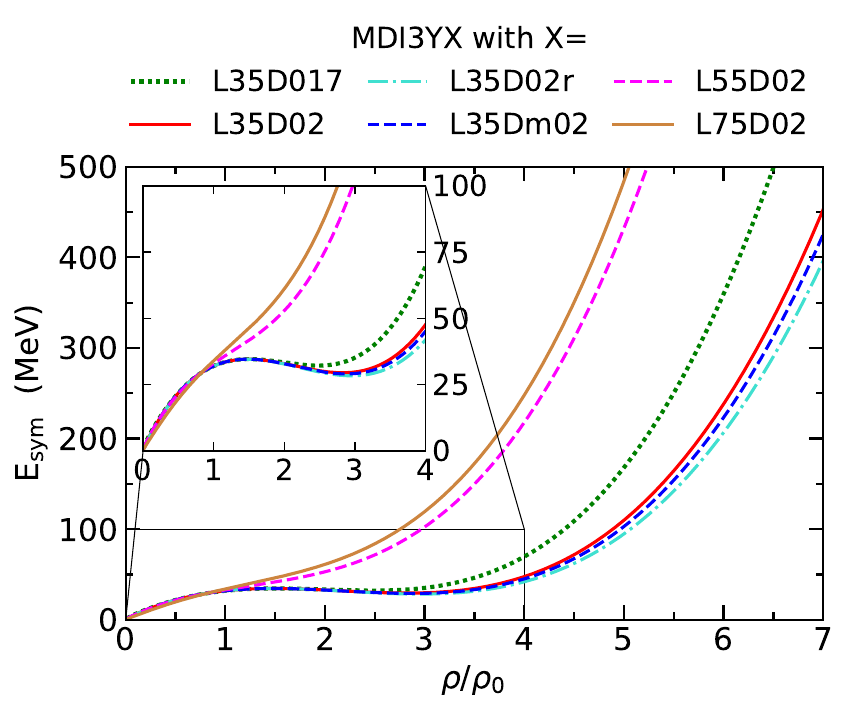}
    \caption{The symmetry energy $E_{{\mathrm{sym}}}(\rho)$ as a function of nucleon density predicted by the MDI3Y interactions.
    }
    \label{fig:Esym}
\end{figure}
%%%%%%%%%%%%%%%%%%%%%%%%%%%%%%%%%%%%%%%%%%%%

Shown in Fig.~\ref{fig:Esym} is the density dependence of the symmetry energy $E_{{\mathrm{sym}}}(\rho)$ obtained with the MDI3Y interactions.
The four interactions with $L=35$~MeV share the same values for $E_{{\mathrm{sym}}}(\rho_0)$, $L$, $K_{{\mathrm{sym}}}$, and $J_{{\mathrm{sym}}}$, which results in similar $E_{{\mathrm{sym}}}(\rho)$ up to approximately $3\rho_0$ except that the MDI3YL35D017 begins to deviate from the others above about $2\rho_0$.
However, due to the different $U_{\mathrm{sym}}$, the $E_{{\mathrm{sym}}}(\rho)$ from the four interactions becomes significantly  different at higher densities, which can also be observed from their different predictions for $I_{{\mathrm{sym}}}$ and $H_{{\mathrm{sym}}}$ in Table~\ref{tab:Macros}.
Furthermore, it is seen that the four interactions with $L=35$~MeV predict a soft $E_{{\mathrm{sym}}}$ at intermediate densities but stiff one at high densities, which could result in a peak structure in the sound speed of neutron star matter~\cite{Wang:2023zcj}.
Moreover, a recent study~\cite{Ye:2024meg} has demonstrated that such a special density behavior of $E_{{\mathrm{sym}}}(\rho)$, which is soft around $2$-$3\rho_0$ but stiff above about $4\rho_0$, can provide a solution to the hyperon puzzle.
It should be noted that the symmetry energy soft around $2$-$3\rho_0$ but stiff above about $4\rho_0$ well agrees with the predictions from the relativistic mean-field (RMF) model with isovector-scalar $\delta$ meson and its coupling to isoscalar-scalar $\sigma$ meson~\citep{Zabari:2018tjk,Zabari:2019clk,Kubis:2020ysv,Miyatsu:2022wuy,Li:2022okx,Miyatsu:2023lki}.
Furthermore, the inclusion of nucleon-nucleon short-range correlations could also soften the $E_{\mathrm{sym}}(\rho)$ at supra-saturation densities in RMF model~\cite{Xu:2009bb,Xu:2012,Cai:2015xga,Burrello:2022tjw}.
In addition, the soft symmetry energy with negative density slope around $2\rho_0$ is also consistent with the constraints~\citep{Xiao:2008vm} obtained from analyzing charged pion ratio in heavy-ion collisions at FOPI~\citep{FOPI:2006ifg} although the results are somewhat model dependent~\citep{Feng:2009am,Xie:2013np,Xu:2013aza,Hong:2013yva,Song:2015hua,Zhang:2017mps,Zhang:2018ool}.
We also note that Liu \textit{et al.}~\cite{Liu:2025pzr} recently simulated the Au+Au collisions at $400$~MeV per nucleon using the UrQMD model, and their analyses of the experimental data for collective flows~\cite{Russotto:2011hq,Russotto:2016ucm} and charged pion productions~\cite{FOPI:2010xrt} support a soft $E_{\mathrm{sym}}(\rho)$ at densities of about $1$-$2\rho_0$.

\subsection{Neutron star properties}
To study the properties of neutron stars, we adopt the conventional neutron star model, in which
the neutron star is assumed to consist of an outer crust, an inner crust, and a uniform core composed of neutrons, protons, electrons and possible muons.
The core-crust transition density $\rho_t$, which separates the liquid core from the nonuniform inner crust, is derived from the so-called thermodynamical method (see, e.g. Ref.~\cite{Xu:2009vi}).
The critical density $\rho_{\mathrm{out}}$ between the inner and the outer crust is taken to be $\rho_{\mathrm{out}}=2.46\times10^{-4}\,\mathrm{fm}^{-3}$~\cite{Carriere:2002bx,Xu:2008vz,Xu:2009vi}.
For the outer crust, where $\rho<\rho_{\mathrm{out}}$, we use the Baym-Pethick-Sutherland EOS~\cite{1971ApJ...170..299B,Iida:1996hh};
for the core, where $\rho>\rho_{t}$, the EOS is obtained using the MDI3Y interactions under the $\beta$-equilibrium and charge-neutralized conditions;
for the inner crust, where $\rho_{\mathrm{out}}<\rho<\rho_t$, we construct the EOS by interpolation with the form~\cite{Carriere:2002bx,Xu:2008vz,Xu:2009vi}
\begin{equation}
\label{eq:innerCrust}
P=a+b \epsilon^{4/3},
\end{equation}
where $P$ and $\epsilon$ represent the pressure and energy density of neutron star matter, respectively, and $a$, $b$ are the coefficients to be determined.
We note that although we employ the polytropic interpolation for inner crust, its EOS remains significantly model-dependent~\cite{Haensel:2007yy,Chamel:2008ca,FiorellaBurgio:2018dga}.
Some more self-consistent methods of treating the inner crust EOS~\cite{Ravenhall:1983uh,Douchin:2001sv,Carreau:2019zdy,Balliet:2020nsh,Thi:2021hai} may reduce the theoretical uncertainties when extracting the nuclear EOS from neutron star observations~\cite{Davis:2024nda,Burrello:2025jay}.

\begin{figure}[ht]
    \centering
    \includegraphics[width=\linewidth]{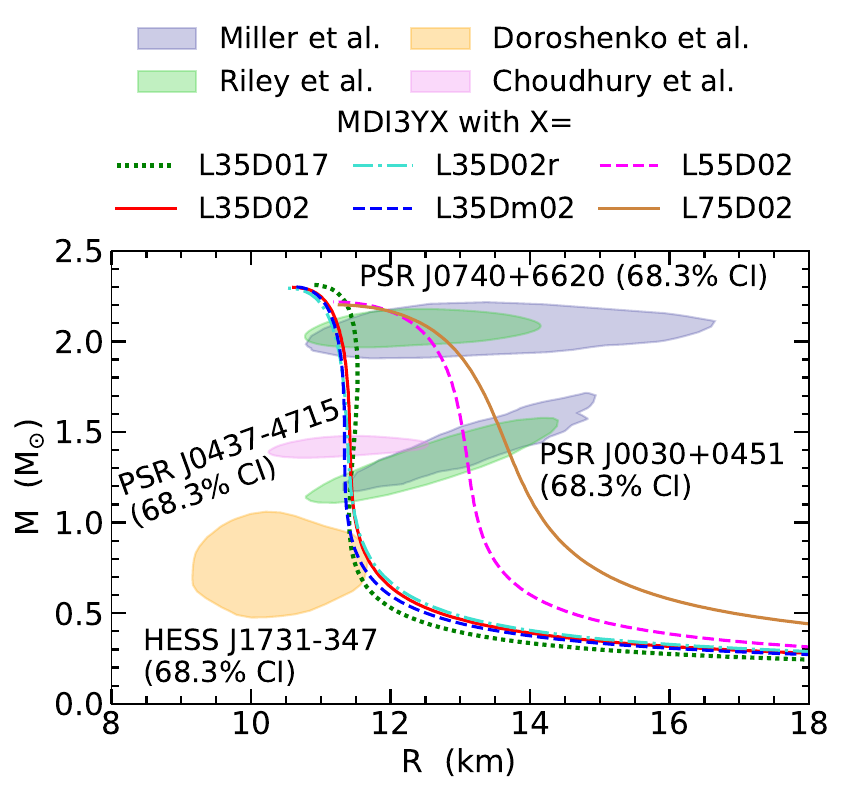}
    \caption{M-R relation for static neutron stars from the MDI3Y interactions.
    The astrophysical constraints for
    PSR J0030+0451 \cite{Miller:2019cac,Riley:2019yda},
    PSR J0437-4715 \cite{Choudhury:2024xbk},
    PSR J0740+6620 \cite{Miller:2021qha,Riley:2021pdl},
    and the CCO in HESS J1731-347 \cite{Doroshenko:2022nwp}
    are also included.
    All contours are plotted for $68.3\%$ CI.
    }
    \label{fig:MR}
\end{figure}

Given the EOS of neutron star matter $P(\epsilon)$, the M-R relation of static neutron stars can be obtained by solving the Tolman-Oppenheimer-Volkoff (TOV) equation~\cite{Tolman:1939jz,Oppenheimer:1939ne}.
Shown in Fig.~\ref{fig:MR} is the M-R relation of neutron stars predicted by the new MDI3Y interactions.
Also shown in Fig.~\ref{fig:MR} are the simultaneous M-R determinations obtained from the observations of various compact stars, including PSR J0030+0451~\cite{Miller:2019cac,Riley:2019yda} and PSR J0437-4715~\cite{Choudhury:2024xbk} with a mass around $1.4$~$M_{\odot}$, PSR J0740+6620~\cite{Miller:2021qha,Riley:2021pdl} with a mass around $2.0$~$M_{\odot}$, as well as the CCO within HESS J1731-347~\cite{Doroshenko:2022nwp} with an unusually low mass around $0.77$~$M_{\odot}$ and small radius around $10.4$~km.
All contours are plotted for $68.3\%$ credible intervals (CI).
It is seen that the four interactions with $L=35$~MeV predict similar M-R relations, and they are all consistent with the measurements of PSR J0030+0451, PSR J0437-4715, PSR J0740+6620, and the CCO in HESS J1731-347 within $68.3\%$ CI, as required in the construction of the four interactions.
These results imply that varying $U_{\mathrm{sym}}$ may have negligible impacts on the maximum mass of neutron stars but noticeable changes on their radii.
In particular,
MDI3YL35D017 predicts smaller radii for low-mass neutron stars but larger radii for massive ones.
As shown in Fig.~\ref{fig:MR}, the M-R relations obtained from MDI3YL55D02 and MDI3YL75D02 are compatible with the observations for both PSR J0030+0451 and PSR J0740+6620, but their radii are too large to meet the constraints for PSR J0437-4715 and the CCO in HESS J1731-347.

%%%%%%%%%%%%%%%%% Figure %%%%%%%%%%%%%%%%%%%
%%%%%%%%%%%%%%%%%%%%%%%%%%%%%%%%%%%%%%%%%%%%

%%%%%%%%%%%%%%%%% Table %%%%%%%%%%%%%%%%%%%%%%%%%%%%%%%%%%%%%%%%%%%%%%%%%%%%%%%%%%%%%%%%%%%%%
\begin{table*}
\caption{\label{tab:MR_data}
    The core-crust transition density $\rho_t$, the radius $R_{1.4}$ and dimensionless tidal deformability $\Lambda_{1.4}$ of $1.4M_{\odot}$ neutron star, as well as the central density $\rho_{\mathrm{cen}}^{\mathrm{TOV}}$ and mass $M_{\mathrm{TOV}}$ of the maximum mass neutron star configuration, predicted by the MDI3YX interactions.
}
\begin{ruledtabular}
\begin{tabular}{ccccccc}
X=& L35D017 & L35D02 & L35D02r& L35Dm02& L55D02& L75D02  \\ \hline
$\rho_t$ ($\mathrm{fm}^{-3}$) &
0.0954 & 0.103  & 0.105 &  0.107  & 0.0880  & 0.0799
\\
$R_{1.4}$ (km) &
11.47 & 11.42 & 11.39 & 11.34 & 13.07  & 13.69
\\
$\Lambda_{1.4}$ &
252.2 & 226.0 & 217.2 & 219.3 & 516.3  & 579.5
\\
$\rho_{\mathrm{cen}}^{\mathrm{TOV}}$ ($\mathrm{fm}^{-3}$) &
0.99 & 1.03 & 1.05 & 1.03 & 1.00  & 1.00
\\
$M_{\mathrm{TOV}}/M_{\odot}$ &
2.31 & 2.30 & 2.29 & 2.30 & 2.22  & 2.20
\\
\end{tabular}
\end{ruledtabular}
\end{table*}
%%%%%%%%%%%%%%%%%%%%%%%%%%%%%%%%%%%%%%%%%%%%%%%%%%%%%%%%%%%%%%%%%%%%%%%%%%%%%%%%%%%%%%

In Table~\ref{tab:MR_data}, we list some neutron star properties, including the core-crust transition density $\rho_t$, the radius $R_{1.4}$ and dimensionless tidal deformability $\Lambda_{1.4}$ of $1.4M_{\odot}$ neutron star, as well as the central density $\rho_{\mathrm{cen}}^{\mathrm{TOV}}$ and mass $M_{\mathrm{TOV}}$ of the maximum mass neutron star configuration.
It is seen from Table~\ref{tab:MR_data} that all the six MDI3Y interactions satisfy the constraint of $\Lambda_{1.4} \le 580$ from the gravitational wave signal GW170817~\cite{LIGOScientific:2018cki}. From the results of the four interactions with $L=35$~MeV, one sees that $U_{\rm sym}$ can have mild influence on the values of $\rho_t$ and $\Lambda_{1.4}$.

\begin{figure}[ht]
    \centering
    \includegraphics[width=\linewidth]{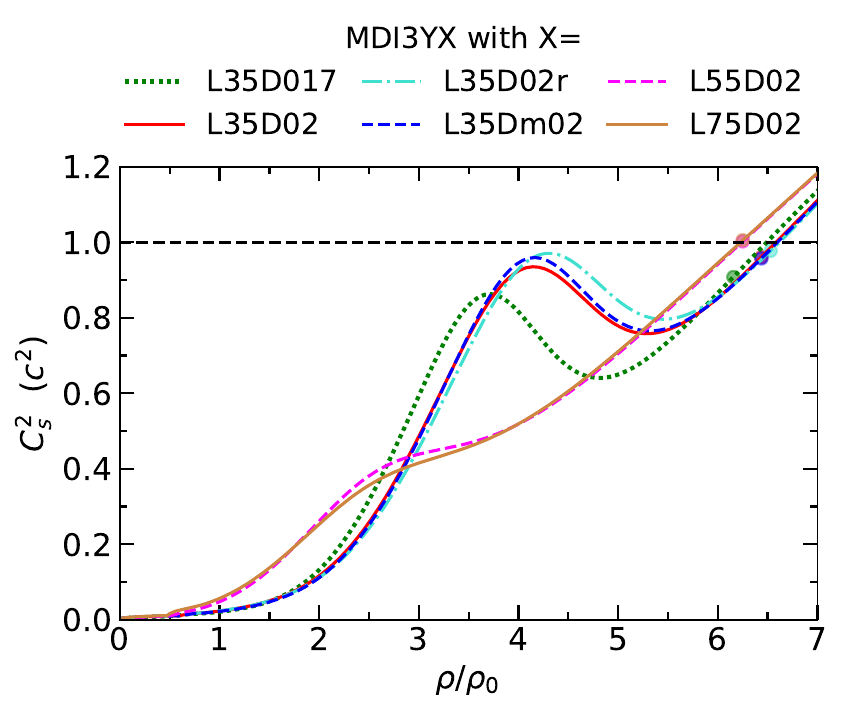}
    \caption{The squared sound speed ($C_{s}^{2}$) of neutron star matter as a function of nucleon density predicted by the MDI3Y interactions.
    The central density ($\rho_{\mathrm{cen}}^{\mathrm{TOV}}$) of maximum mass neutron star configuration is marked as solid symbol.}
    \label{fig:Cs2}
\end{figure}

Shown in Fig.~\ref{fig:Cs2} is the squared sound speed $C_{s}^{2}\equiv \mathrm{d}P/\mathrm{d}\epsilon$ of neutron star matter as a function of nucleon density, and the location of the $\rho_{\mathrm{cen}}^{\mathrm{TOV}}$ is marked.
It is seen that the causality condition $C_{s}^{2}\leq c^2 $ is satisfied for all the six interactions.
An obvious peak structure can be observed in the four interactions with $L=35$~MeV, which results from the corresponding symmetry energy that is soft at intermediate densities but stiff at high densities by construction, as shown in Fig.~\ref{fig:Esym}, and varying $U_{\mathrm{sym}}$ can change the location and height of the peak.
The peak disappears when $L$ increases to $55$ and $75$~MeV, which is consistent with the conclusions in Ref.~\cite{Wang:2023zcj}.

%%%%%%%%%%%%%%%%% Figure %%%%%%%%%%%%%%%%%%%
%%%%%%%%%%%%%%%%%%%%%%%%%%%%%%%%%%%%%%%%%%%%

\section{SUMMARY AND OUTLOOK}
\label{sec:summary}
We have extended the traditional MDI model to the so-called MDI3Y model by including two additional finite-range Yukawa interactions and modifying the DD terms following the idea of Fermi momentum expansion.
This extension significantly enhances the flexibility of momentum behavior of the single-nucleon potential.
For example, the single-nucleon potential can have very different high energy behaviors while nicely describe the known empirical data at lower energies; the isoscalar nucleon effective mass can exhibit a peak structure in the density dependence with $m_{s}^{\ast}/m > 1$ at low densities;
and the symmetry potential can display non-monotonic momentum dependence by independently adjusting the values of the linear isospin splitting coefficient $\Delta m_{1}^{\ast}(\rho)$ of nucleon effective mass, the isovector nucleon effective mass $m_{v}^{\ast}$ and the limit of the symmetry potential at infinite nucleon momentum $U_{\mathrm{sym}}^{\infty}$.
At the same time, due to the extended DD terms, the MDI3Y model exhibits highly flexible density dependence of nuclear matter EOS, especially for the symmetry energy.

Based on the MDI3Y model, we have built a series of interactions with various symmetry potentials, including both monotonic and non-monotonic, and different symmetry energy behaviors.
The parameters of these interactions are obtained under the constraints from microscopic calculations, terrestrial experiments and astrophysical observations.
With these new interactions, we have studied the properties of nuclear matter and neutron stars.
Moreover, the exchange contribution from the finite-range interaction can be approximated by that from a Skyrme-like zero-range interaction based on density matrix expansion, and the corresponding energy density and single-nucleon potential are expressed in terms of local densities.

In future, the MDI3Y interactions constructed in the present work can be used to replace the traditional MDI interactions in transport models, allowing for a more comprehensive investigation of the symmetry energy, symmetry potential, and the isospin splitting of nucleon effective masses from heavy-ion collisions.
In addition, the MDI3Y model can be further extended to include octet baryons to study the properties of hyperon stars and the hyperon production in heavy-ion collisions.

\begin{acknowledgments}
We are grateful to Xin Li, Rui Wang, Jun-Ting Ye, Ying-Xun Zhang, and Zhen Zhang for useful discussions.
This work was supported in part by the National Natural Science Foundation of China under Grant No 12235010, the National SKA Program of China (Grant No. 2020SKA0120300), and the Science and Technology Commission of Shanghai Municipality (Grant No. 23JC1402700).
\end{acknowledgments}

\appendix

\section{RELATIONS BETWEEN PARAMETERS IN MDI3Y AND M3Y-P$n$ INTERACTIONS}
\label{sec:Nakada}
The central part of the two-body interaction in M3Y-P$n$~\cite{Nakada:2003fw,Nakada:2008eh,Nakada:2009xh,Nakada:2012sq} is written as
\begin{equation}
\begin{aligned}
\label{eq:Nakada_C}
V_{\mathrm{M3Y}}^{C}(\vec{r}_1,\vec{r}_2) =& \sum_{i=1}^{3}  \bigg[   t_{i}^{(\mathrm{SE})}P_{\mathrm{SE}} + t_{i}^{(\mathrm{TE})}P_{\mathrm{TE}} \\ &+ t_{i}^{(\mathrm{SO})}P_{\mathrm{SO}} + t_{i}^{(\mathrm{TO})}P_{\mathrm{TO}}\bigg]
\frac{e^{-\mu_{i}\left|\vec{r}_1-\vec{r}_2\right|}}{\mu_{i}\left|\vec{r}_1-\vec{r}_2\right|},
\end{aligned}
\end{equation}
where the projection operators on the singlet-even (SE), triplet-even (TE), singlet-odd (SO), and triplet-odd (TO) states are
\begin{equation}
\label{eq:STproj}
\begin{aligned}
P_{\mathrm{SE}}&=\frac{1-P_\sigma}{2}\frac{1+P_\tau}{2},~~
P_{\mathrm{TE}}=\frac{1+P_\sigma}{2}\frac{1-P_\tau}{2},  \\
P_{\mathrm{SO}}&=\frac{1-P_\sigma}{2}\frac{1-P_\tau}{2},~~
P_{\mathrm{TO}}=\frac{1+P_\sigma}{2}\frac{1+P_\tau}{2}. \\
\end{aligned}
\end{equation}
By substituting Eq.~(\ref{eq:STproj}) into Eq.~(\ref{eq:Nakada_C}) and comparing with Eq.~(\ref{eq:v3Y}), we can obtain the relations between the parameters as~\cite{Nakada:2003fw}
\begin{equation}
\label{eq:C_rlts}
\begin{aligned}
t_{i}^{(\mathrm{SE})} =& W_i -B_i -H_i +M_i , \\
t_{i}^{(\mathrm{TE})} =& W_i +B_i +H_i +M_i ,\\
t_{i}^{(\mathrm{SO})} =& W_i -B_i +H_i -M_i ,\\
t_{i}^{(\mathrm{TO})} =& W_i +B_i -H_i -M_i .
\end{aligned}
\end{equation}
The first version of M3Y-P$n$ interactions~\cite{Nakada:2003fw} introduced the DD term as Eq.~(\ref{eq:V_DD}) with $n=1$.
In the subsequent versions of M3Y-P$n$~\cite{Nakada:2008eh,Nakada:2009xh,Nakada:2012sq}, the DD term was replaced by
\begin{equation}
\label{eq:Nakada_DD}
\begin{aligned}
V_{\mathrm{M3Y}}^{\mathrm{DD}}=& \Big(t_{\rho}^{(\mathrm{SE})} P_{\mathrm{SE}}[\rho(\vec{R})]^{\alpha^{(\mathrm{SE})}}  \\
& +  t_{\rho}^{(\mathrm{TE})} P_{\mathrm{TE}}[\rho(\vec{R})]^{\alpha^{(\mathrm{TE})}} \Big)
\delta(\vec{r}_{1}-\vec{r}_{2}),
\end{aligned}
\end{equation}
with $\alpha^{(\mathrm{TE})}=1/3$ and $\alpha^{(\mathrm{SE})}=1$.
The Eq.~(\ref{eq:V_DD}) can be rewritten as
\begin{equation}
\label{eq:V_DD_ST}
\begin{aligned}
V^{\mathrm{DD}}=&\sum_{n=1,3,5} \frac{1}{6}t_{3}^{[n]} \Big[
(1-x_{3}^{[n]})(P_{\mathrm{SE}}+P_{\mathrm{SO}}) \\
&+ (1+x_{3}^{[n]})(P_{\mathrm{TE}}+P_{\mathrm{TO}}) \Big] \rho^{n/3}(\vec{R}) \delta(\vec{r}_{1}-\vec{r}_{2}).
\end{aligned}
\end{equation}
Considering that $V^{\mathrm{DD}}$ has no contributions to the SO and TO channels when calculating Eq.~(\ref{eq:E_pot}), Eq.~(\ref{eq:V_DD_ST}) will reduce to Eq.~(\ref{eq:Nakada_DD}) when taking $x_{3}^{[1]}=1$, $x_{3}^{[3]}=-1$, and $t_{3}^{[5]}=0$.
We then obtain $t_{\rho}^{(\mathrm{TE})} = \frac{1}{3} t_{3}^{[1]}$ and $t_{\rho}^{(\mathrm{SE})} = \frac{1}{3} t_{3}^{[3]}$.

\section{HARTREE-FOCK CALCULATIONS WITH FINITE-RANGE INTERACTION}
\label{sec:HF_DME}
\subsection{HAMILTONIAN DENSITY}
\label{sec:Vc_HF}
The contribution from the finite-range interaction $V^C$ in Eq.~(\ref{eq:E_pot}) can be obtained as
\begin{equation}
\label{eq:E_pot_Vc}
\mathrm{H}_r =  \mathrm{H}_{r}^{D} + \mathrm{H}_{r}^{E},
\end{equation}
where
\begin{equation}
\small
\label{eq:E_pot_Vc_D}
\begin{aligned}
\mathrm{H}_{r}^{D} = & \frac{1}{2} \sum_{i,j} \langle i j| V^{C} |ij\rangle \\
=& \frac{1}{2} \sum_{i=1}^{3} \int \! \! \! \! \int \mathrm{d}^3 r_1 \mathrm{d}^3 r_2
\frac{e^{-\mu_i |\vec{r}_1 - \vec{r}_2|}}{\mu_i |\vec{r}_1 - \vec{r}_2|}\bigg\{ (W_i + \frac{B_i}{2}-H_i-\frac{M_i}{2}) \\
&\times  \Big[ \rho_{\mathrm{n}}(\vec{r}_1) \rho_{\mathrm{n}}(\vec{r}_2) + \rho_{\mathrm{p}}(\vec{r}_1) \rho_{\mathrm{p}}(\vec{r}_2)\Big] \\
&+ 2 (W_i+\frac{B_i}{2}) \rho_{\mathrm{n}}(\vec{r}_1) \rho_{\mathrm{p}}(\vec{r}_2)
\bigg\},
\end{aligned}
\end{equation}
and
\begin{equation}
\small
\label{eq:E_pot_Vc_E}
\begin{aligned}
\mathrm{H}_{r}^{E} = & \frac{1}{2} \sum_{i,j} \langle i j| V^{C} (-P_M P_\sigma P_\tau) |ij\rangle \\
=& \frac{1}{2} \sum_{i=1}^{3} \int \! \! \! \! \int \mathrm{d}^3 r_1 \mathrm{d}^3 r_2
\frac{e^{-\mu_i |\vec{r}_1 - \vec{r}_2|}}{\mu_i |\vec{r}_1 - \vec{r}_2|}\bigg\{ (M_i + \frac{H_i}{2}-B_i-\frac{W_i}{2}) \\
&\times  \Big[ \rho_{\mathrm{n}}(\vec{r}_1,\vec{r}_2) \rho_{\mathrm{n}}(\vec{r}_2,\vec{r}_1) + \rho_{\mathrm{p}}(\vec{r}_1,\vec{r}_2)
\rho_{\mathrm{p}}(\vec{r}_2,\vec{r}_1)\Big] \\
&+ 2 (M_i+\frac{H_i}{2}) \rho_{\mathrm{n}}(\vec{r}_1,\vec{r}_2)
\rho_{\mathrm{p}}(\vec{r}_2,\vec{r}_1)
\bigg\},
\end{aligned}
\end{equation}
represent the direct and exchange contributions, respectively.
In the derivations of Eqs.~(\ref{eq:E_pot_Vc_D}) and (\ref{eq:E_pot_Vc_E}), $P_\tau$ is replaced by $\delta_{\tau_1 \tau_2}$ since there is no isospin mixing state, and the effect of $P_\sigma$ is simply replaced by a factor of $\frac{1}{2}$ for the spin-averaged quantities.
The matrix element of the density operator $\hat \rho_{\tau} = \sum_i | i \tau\rangle \langle i \tau|$ is expressed as $\rho_{\tau}(\vec{r}_1,\vec{r}_2)= \langle \vec{r}_1 | \hat \rho_{\tau} | \vec{r}_2 \rangle $, which reduces to the local density when $\vec{r}_1 = \vec{r}_2$.

To obtain the energy density functional from finite-range interaction, we introduce the coordinate transformation:
\begin{equation}
\label{eq:coord_trans}
\vec{r}=(\vec{r}_{1}+\vec{r}_{2})/2, ~~
\vec{s}=\vec{r}_{1}-\vec{r}_{2}.
\end{equation}
The direct contribution can be expressed as
\begin{equation}
\mathrm{H}_{r}^{D} = \int \mathrm{d}^3 r \mathcal{H}_{r}^{D}(\vec{r}),
\end{equation}
where
\begin{equation}
\label{eq:Hdr_general}
\begin{aligned}
\mathcal{H}_{r}^{D}(\vec{r})
=& \frac{1}{2} \sum_{i=1}^{3} \int \mathrm{d}^3 s
\frac{e^{-\mu_i s}}{\mu_i s}\bigg\{ (W_i + \frac{B_i}{2}-H_i-\frac{M_i}{2}) \\
&\times  \Big[ \rho_{\mathrm{n}}(\vec{r}+ \frac{\vec{s}}{2}) \rho_{\mathrm{n}}(\vec{r}- \frac{\vec{s}}{2}) + \rho_{\mathrm{p}}(\vec{r}+ \frac{\vec{s}}{2}) \rho_{\mathrm{p}}(\vec{r}- \frac{\vec{s}}{2})\Big] \\
&+ 2 (W_i+\frac{B_i}{2}) \rho_{\mathrm{n}}(\vec{r}+ \frac{\vec{s}}{2}) \rho_{\mathrm{p}}(\vec{r}- \frac{\vec{s}}{2})
\bigg\}.
\end{aligned}
\end{equation}
With the integral $\int \mathrm{d}^3 s e^{-\mu s}/s = 4 \pi/\mu^2 $, direct contribution reduces to the Skyrme type for uniform nuclear matter:
\begin{equation}
\label{eq:Hdr_general_NM}
\begin{aligned}
\mathcal{H}_{r}^{D}(\vec{r}) =& \bigg[ \sum_{i=1}^{3} \frac{4\pi}{\mu_{i}^{3}} (W_i + \frac{B_i}{2} - H_i-\frac{M_i}{2})\bigg]
\frac{\rho_{\mathrm{n}}^{2}+\rho_{\mathrm{p}}^{2}}{2} \\
&+ \bigg[ \sum_{i=1}^{3} \frac{4\pi}{\mu_{i}^{3}} (W_i + \frac{B_i}{2}) \bigg] \rho_{\mathrm{n}} \rho_{\mathrm{p}},
\end{aligned}
\end{equation}
which is denoted as $\mathcal{H}^{\mathrm{loc}}(\vec{r})$ [Eq.~(\ref{eq:H_loc})] in the main text.

Expressing the density matrix in terms of the Wigner function through the Fourier transform:
\begin{equation}
\begin{aligned}
&f_{\tau}(\vec{r},\vec{p}) = \frac{1}{(2\pi\hbar)^3} \int
\rho_{\tau}(\vec{r}+ \frac{\vec{s}}{2},\vec{r}- \frac{\vec{s}}{2})
e^{- \mathrm{i} \vec{p} \cdot \vec{s}/\hbar} \mathrm{d}^3 s, \\
&\rho_{\tau}(\vec{r}+ \frac{\vec{s}}{2},\vec{r}- \frac{\vec{s}}{2}) = \int f_{\tau}(\vec{r},\vec{p}) e^{\mathrm{i}  \vec{p} \cdot \vec{s}/\hbar}  \mathrm{d}^3 p,
\end{aligned}
\end{equation}
the integrals in the exchange contribution can be rewritten as
\begin{equation}
\begin{aligned}
& \int \mathrm{d}^3r \mathrm{d}^3s \frac{e^{-\mu s}}{\mu s}
\rho_{\tau}(\vec{r}+ \frac{\vec{s}}{2},\vec{r}- \frac{\vec{s}}{2})
\rho_{\tau^\prime}(\vec{r}- \frac{\vec{s}}{2},\vec{r}+ \frac{\vec{s}}{2}) \\
=& \int \mathrm{d}^3r  \mathrm{d}^3 p \mathrm{d}^3 p^{\prime} \mathrm{d}^3s\frac{e^{-\mu s}}{\mu s}
f_{\tau}(\vec{r},\vec{p}) e^{\mathrm{i}  \vec{p} \cdot \vec{s}/\hbar}
f_{\tau^\prime}(\vec{r},\vec{p}^{\, \prime}) e^{-\mathrm{i} \vec{p}^{\, \prime} \cdot \vec{s}/\hbar} \\
=& \int \mathrm{d}^3r \mathrm{d}^3 p \mathrm{d}^3 p^{\prime}
\frac{4 \pi}{\mu^3} \frac{f_{\tau}(\vec{r},\vec{p}) f_{\tau^\prime}(\vec{r},\vec{p}^{\, \prime}) }{1+\left(\vec{p}-\vec{p}^{\, \prime}\right)^2 / (\hbar \mu)^2},
\end{aligned}
\end{equation}
where we have used the integral
\begin{equation}
\int \mathrm{d}^3 s e^{\mathrm{i} (\vec{p}-\vec{p}^{\, \prime})\cdot \vec{s}/\hbar} \frac{e^{-\mu s}}{\mu s} = \frac{4 \pi}{\mu^3}
\frac{1}{1+\left(\vec{p}-\vec{p}^{\, \prime}\right)^2 / (\hbar \mu)^2}.
\end{equation}
The exchange contribution can then be expressed as
\begin{equation}
\mathrm{H}_{r}^{E} = \int \mathrm{d}^3 r \mathcal{H}_{r}^{E}(\vec{r}),
\end{equation}
where
\begin{equation}
\label{eq:Her_f}
\begin{aligned}
\mathcal{H}_{r}^{E}(\vec{r}) =& \frac{1}{2} \sum_{i=1}^{3} \frac{4 \pi}{\mu_{i}^{3}} \int \! \! \! \! \int \frac{\mathrm{d}^3 p \mathrm{~d}^3 p^{\prime}}{1+\left(\vec{p}-\vec{p}^{\, \prime}\right)^2 / (\hbar \mu_i)^2} \\
&\times \bigg\{(M_i + \frac{H_i}{2} -B_i -\frac{W_i}{2}) \\
& \times \Big[ f_{\mathrm{n}}(\vec{r}, \vec{p}) f_{\mathrm{n}}\left(\vec{r}, \vec{p}^{\, \prime}\right) + f_{\mathrm{p}}(\vec{r}, \vec{p}) f_{\mathrm{p}}\left(\vec{r}, \vec{p}^{\, \prime}\right)  \Big] \\
&+2(M_i + \frac{H_i}{2}) f_{\mathrm{n}}(\vec{r}, \vec{p}) f_{\mathrm{p}}\left(\vec{r}, \vec{p}^{\, \prime}\right) \bigg\},
\end{aligned}
\end{equation}
which is denoted as $\mathcal{H}^{\mathrm{MD}}(\vec{r})$ [Eq.~(\ref{eq:H_MD})] in the main text.

\subsection{DENSITY MATRIX EXPANSION}
\label{sec:DME}
Based on the density matrix expansion~\cite{Negele:1972zp} and by analogy with the results for the MDI model~\cite{Xu:2010ce}, the exchange energy density $\mathcal{H}_{r}^{E}(\vec{r})$ from finite-range interaction can be approximated by the Skyrme-like zero-range energy density $\mathcal{H}_{\mathrm{SL}}^{E}(\vec{r})$, which is expressed in terms of densities $\rho_{\tau}(\vec{r})$ as well as kinetic energy densities $\uptau_{\tau}(\vec{{r}})=\sum_{i}|\bm{\nabla} \langle \vec{r} | i \tau\rangle|^{2}$:
\begin{equation}
\label{eq:H_SL}
\begin{aligned}
\mathcal{H}_{\mathrm{SL}}^{E}(\vec{r}) = &
A[\rho_{\mathrm{n}}(\vec{r}),\rho_{\mathrm{p}}(\vec{r})] \\
&+B[\rho_{\mathrm{n}}(\vec{r}),\rho_{\mathrm{p}}(\vec{r})]\uptau_{\mathrm{n}}(\vec{{r}})
+B[\rho_{\mathrm{p}}(\vec{r}),\rho_{\mathrm{n}}(\vec{r})]\uptau_{\mathrm{p}}(\vec{{r}}) \\
&+C[\rho_{\mathrm{n}}(\vec{r}),\rho_{\mathrm{p}}(\vec{r})] |\bm{\nabla} \rho_{\mathrm{n}}(\vec{r})|^{2} \\
&+C[\rho_{\mathrm{p}}(\vec{r}),\rho_{\mathrm{n}}(\vec{r})] |\bm{\nabla} \rho_{\mathrm{p}}(\vec{r})|^{2} \\
&+D[\rho_{\mathrm{p}}(\vec{r}),\rho_{\mathrm{n}}(\vec{r})] \bm{\nabla} \rho_{\mathrm{n}}(\vec{r}) \cdot \bm{\nabla} \rho_{\mathrm{p}}(\vec{r}),
\end{aligned}
\end{equation}
where
\begin{align}
A(\rho_{\mathrm{n}},\rho_{\mathrm{p}}) =& V_{\mathrm{NM}}(\rho_{\mathrm{n}},\rho_{\mathrm{p}}) - \frac{3}{5} (3\pi^2)^{2/3} [\rho_{\mathrm{n}}^{5/3} B(\rho_{\mathrm{n}},\rho_{\mathrm{p}}) \nonumber \\
& + \rho_{\mathrm{p}}^{5/3} B(\rho_{\mathrm{p}},\rho_{\mathrm{n}})], \\
B(\rho_{\mathrm{n}},\rho_{\mathrm{p}}) =& -2[\rho_{\mathrm{n}} V^{L}(\rho_{\mathrm{n}})+\rho_{\mathrm{p}}V^{U}(\rho_{\mathrm{p}},\rho_{\mathrm{n}})], \\
C(\rho_{\mathrm{n}},\rho_{\mathrm{p}}) =& -\frac{\partial F(\rho_{\mathrm{n}},\rho_{\mathrm{p}})}{\partial \rho_{\mathrm{n}}}, \\
D(\rho_{\mathrm{n}},\rho_{\mathrm{p}}) =& -\frac{\partial F(\rho_{\mathrm{n}},\rho_{\mathrm{p}})}{\partial \rho_{\mathrm{p}}} - \frac{\partial F(\rho_{\mathrm{p}},\rho_{\mathrm{n}})}{\partial \rho_{\mathrm{n}}}.
\end{align}
We have defined that
\begin{equation}
\begin{aligned}
V_{\mathrm{NM}}(\rho_{\mathrm{n}},\rho_{\mathrm{p}}) =& \rho_{\mathrm{n}}^{2} V_{\mathrm{NM}}^{L}(\rho_{\mathrm{n}}) + \rho_{\mathrm{p}}^{2} V_{\mathrm{NM}}^{L}(\rho_{\mathrm{p}}) \\
&+ 2 \rho_{\mathrm{n}} \rho_{\mathrm{p}} V_{\mathrm{NM}}^{U}(\rho_{\mathrm{n}},\rho_{\mathrm{p}}),
\end{aligned}
\end{equation}
with
\begin{equation}
\begin{aligned}
V_{\mathrm{NM}}^{L}(\rho_\tau) =& \frac{1}{2} \sum_{i=1}^{3} (M_i+\frac{H_i}{2}-B_i-\frac{W_i}{2} ) \\
&\times \int \mathrm{d}^3 s \rho_{\mathrm{SL}}^{2}(k_{\tau} s) \frac{e^{- \mu_i s}}{\mu_i s},
\end{aligned}
\end{equation}
\begin{equation}
\begin{aligned}
V_{\mathrm{NM}}^{U}(\rho_\tau,\rho_{\tau^{\prime}}) =& \frac{1}{2} \sum_{i=1}^{3} (M_i+\frac{H_i}{2} ) \\
&\times \int \mathrm{d}^3 s \rho_{\mathrm{SL}}(k_{\tau} s) \rho_{\mathrm{SL}}(k_{\tau^{\prime}} s) \frac{e^{- \mu_i s}}{\mu_i s}.
\end{aligned}
\end{equation}
Other terms are defined as
\begin{equation}
F(\rho_\tau,\rho_{\tau^{\prime}}) =
\frac{1}{2} V^L(\rho_\tau) \rho_\tau  + \frac{1}{2} V^U(\rho_\tau,\rho_{\tau^{\prime}}) \rho_{\tau^{\prime}},
\end{equation}
\begin{equation}
\begin{aligned}
V^L(\rho_\tau) = & \frac{1}{2} \sum_{i=1}^{3} (M_i+\frac{H_i}{2}-B_i-\frac{W_i}{2} ) \\
& \times \int \mathrm{d}^3 s s^2 \rho_{\mathrm{SL}}(k_{\tau} s) g(k_{\tau} s) \frac{e^{- \mu_i s}}{\mu_i s},
\end{aligned}
\end{equation}
\begin{equation}
\begin{aligned}
V^U(\rho_\tau,\rho_{\tau^\prime}) =& \frac{1}{2} \sum_{i=1}^{3} (M_i + \frac{H_i}{2}) \\
& \times \int \mathrm{d}^3 s s^2 \rho_{\mathrm{SL}}(k_{\tau} s) g(k_{\tau^\prime} s) \frac{e^{- \mu_i s}}{\mu_i s},
\end{aligned}
\end{equation}
where
\begin{align}
\rho_{\mathrm{SL}}(k_{\tau} s) =& \frac{3}{k_{\tau} s} j_1 (k_{\tau} s),\\
g(k_{\tau} s) =& \frac{35}{2(k_{\tau} s)^3} j_3 (k_{\tau} s),
\end{align}
with $j_1$ and $j_3$ being, respectively, the first- and third-order spherical Bessel functions and $k_\tau=(3\pi^2 \rho_\tau)^{1/3}$ being the Fermi wave number.
It is seen that $A(\rho_{\mathrm{n}},\rho_{\mathrm{p}})$ and $D(\rho_{\mathrm{n}},\rho_{\mathrm{p}})$ are symmetric in $\rho_{\mathrm{n}}$ and $\rho_{\mathrm{p}}$, while $B(\rho_{\mathrm{n}},\rho_{\mathrm{p}})$ and $C(\rho_{\mathrm{n}},\rho_{\mathrm{p}})$ are not.

\begin{widetext}
The single nucleon potential from finite-range interaction can be expressed as
\begin{equation}
U_{r,\tau}(\rho_{\mathrm{n}},\rho_{\mathrm{p}}) = U_{r,\tau}^{D}(\rho_{\mathrm{n}},\rho_{\mathrm{p}}) + U_{r,\tau}^{E}(\rho_{\mathrm{n}},\rho_{\mathrm{p}}),
\end{equation}
with the direct contribution being
\begin{equation}
\begin{aligned}
U_{r,\tau}^{D}(\rho_{\mathrm{n}},\rho_{\mathrm{p}}) = \frac{\partial \mathcal{H}_{r}^{D}(\rho_{\mathrm{n}},\rho_{\mathrm{p}})}{\partial \rho_\tau}
=\sum_{i=1}^{3}
\int \mathrm{d}^3 r^\prime \frac{e^{- \mu_i |\vec{r}-\vec{r}^{\, \prime}|}}{\mu_i |\vec{r}-\vec{r}^{\, \prime}|} \bigg[
(W_i +\frac{B_i}{2}-H_i-\frac{M_i}{2}) \rho_{\tau}(\vec{r}^{\, \prime})
+ (W_i +\frac{B_i}{2}) \rho_{-\tau}(\vec{r}^{\, \prime})
\bigg] ,
\end{aligned}
\end{equation}
which reduces to $(A_l \rho_\tau + A_u \rho_{-\tau})/\rho_0 $ in the uniform nuclear matter, as given in Eq.~(\ref{eq:Utau}).
Since $\mathcal{H}_{\mathrm{SL}}^{E}(\vec{r})$ includes terms with density gradients, the exchange contribution can be calculated as~\cite{Kolomietz:2017qkb}
\begin{equation}
\begin{aligned}
U_{r,\tau}^{E}(\rho_{\mathrm{n}},\rho_{\mathrm{p}}) =& \frac{\partial \mathcal{H}_{\mathrm{SL}}^{E}(\rho_{\mathrm{n}},\rho_{\mathrm{p}})}{\partial \rho_\tau} - \bm{\nabla} \cdot \frac{\partial \mathcal{H}_{\mathrm{SL}}^{E}(\rho_{\mathrm{n}},\rho_{\mathrm{p}})}{\partial ( \bm{\nabla} \rho_\tau)} \\
=& \frac{\partial A(\rho_{\mathrm{n}},\rho_{\mathrm{p}})}{\partial \rho_\tau}
+ \frac{\partial B(\rho_{\mathrm{n}},\rho_{\mathrm{p}})}{\partial \rho_\tau} \uptau_{\mathrm{n}}
+ \frac{\partial B(\rho_{\mathrm{p}},\rho_{\mathrm{n}})}{\partial \rho_\tau} \uptau_{\mathrm{p}} \\
&+ \frac{\partial C(\rho_{\mathrm{n}},\rho_{\mathrm{p}})}{\partial \rho_\tau} (\bm{\nabla} \rho_{\mathrm{n}})^2
+ \frac{\partial C(\rho_{\mathrm{p}},\rho_{\mathrm{n}})}{\partial \rho_\tau} (\bm{\nabla} \rho_{\mathrm{p}})^2
+ \frac{\partial D(\rho_{\mathrm{n}},\rho_{\mathrm{p}})}{\partial \rho_\tau}  \bm{\nabla} \rho_{\mathrm{n}} \cdot \bm{\nabla} \rho_{\mathrm{p}} \\
&-2 \delta_{\tau,n} \bigg[ \frac{\partial C(\rho_{\mathrm{n}},\rho_{\mathrm{p}})}{ \partial \rho_{\mathrm{n}}} (\bm{\nabla} \rho_{\mathrm{n}})^2
+ \frac{\partial C(\rho_{\mathrm{n}},\rho_{\mathrm{p}})}{ \partial \rho_{\mathrm{p}}} \bm{\nabla} \rho_{\mathrm{n}} \cdot \bm{\nabla} \rho_{\mathrm{p}}
+ C(\rho_{\mathrm{n}},\rho_{\mathrm{p}}) \nabla^2 \rho_{\mathrm{n}}
\bigg] \\
&-2 \delta_{\tau,p} \bigg[ \frac{\partial C(\rho_{\mathrm{p}},\rho_{\mathrm{n}})}{ \partial \rho_{\mathrm{p}}}  (\bm{\nabla} \rho_{\mathrm{p}})^2
+ \frac{\partial C(\rho_{\mathrm{p}},\rho_{\mathrm{n}})}{ \partial \rho_{\mathrm{n}}} \bm{\nabla} \rho_{\mathrm{n}} \cdot \bm{\nabla} \rho_{\mathrm{p}}
+ C(\rho_{\mathrm{p}},\rho_{\mathrm{n}}) \nabla^2 \rho_{\mathrm{p}}
\bigg] \\
&- \bigg[ \frac{\partial D(\rho_{\mathrm{n}},\rho_{\mathrm{p}})}{ \partial \rho_{\mathrm{n}}} \bm{\nabla} \rho_{\mathrm{n}} \cdot \bm{\nabla} \rho_{-\tau}
+ \frac{\partial D(\rho_{\mathrm{n}},\rho_{\mathrm{p}})}{ \partial \rho_{\mathrm{p}}} \bm{\nabla} \rho_{\mathrm{p}} \cdot \bm{\nabla} \rho_{-\tau}
+ D(\rho_{\mathrm{n}},\rho_{\mathrm{p}}) \nabla^2 \rho_{-\tau}
\bigg].
\end{aligned}
\end{equation}
With the extended Thomas-Fermi approximation~\cite{Brack:1985vp,Jen76}:
\begin{equation}
\uptau_\tau = a \rho_{\tau}^{5/3} + b\frac{(\bm{\nabla} \rho_\tau)^2}{\rho_\tau} + c \nabla^2 \rho_\tau,
\end{equation}
where $a=\frac{3}{5}(3\pi^2)^{2/3}$, $b=1/36$, and $c=1/3$, the single-nucleon potential from finite-range interaction can then be written as
\begin{equation}
U_{r,\tau}(\rho_{\mathrm{n}},\rho_{\mathrm{p}}) = U_{r,\tau}^{D}(\rho_{\mathrm{n}},\rho_{\mathrm{p}}) + U_{r,\tau}^{0}(\rho_{\mathrm{n}},\rho_{\mathrm{p}}) + U_{r,\tau}^{\nabla}(\rho_{\mathrm{n}},\rho_{\mathrm{p}}),
\end{equation}
where
\begin{equation}
U_{r,\tau}^{0}(\rho_{\mathrm{n}},\rho_{\mathrm{p}}) =
\frac{\partial A(\rho_{\mathrm{n}},\rho_{\mathrm{p}})}{\partial \rho_\tau}
+ a \frac{\partial B(\rho_{\mathrm{n}},\rho_{\mathrm{p}})}{\partial \rho_\tau} \rho_{\mathrm{n}}^{5/3}
+ a \frac{\partial B(\rho_{\mathrm{p}},\rho_{\mathrm{n}})}{\partial \rho_\tau} \rho_{\mathrm{p}}^{5/3},
\end{equation}
\begin{equation}
\begin{aligned}
U_{r,\tau}^{\nabla}(\rho_{\mathrm{n}},\rho_{\mathrm{p}}) =
G_{\tau}^{\mathrm{nn}1} (\bm{\nabla}\rho_{\mathrm{n}})^2
+ G_{\tau}^{\mathrm{pp}1} (\bm{\nabla}\rho_{\mathrm{p}})^2
+ G_{\tau}^{\mathrm{np}} \bm{\nabla}\rho_{\mathrm{n}} \cdot \bm{\nabla}\rho_{\mathrm{p}}
+ G_{\tau}^{\mathrm{nn}2} \nabla^2 \rho_{\mathrm{n}}
+ G_{\tau}^{\mathrm{pp}2} \nabla^2 \rho_{\mathrm{p}},
\end{aligned}
\end{equation}
with
\begin{equation}
G_{\tau}^{\mathrm{nn}1} = \frac{b}{\rho_{\mathrm{n}}} \frac{\partial B(\rho_{\mathrm{n}},\rho_{\mathrm{p}})}{\partial \rho_\tau}
+ (1-2\delta_{\tau,\mathrm{n}}) \frac{\partial C(\rho_{\mathrm{n}},\rho_{\mathrm{p}})}{\partial \rho_\tau}
- \delta_{\tau,\mathrm{p}} \frac{\partial D(\rho_{\mathrm{n}},\rho_{\mathrm{p}})}{\partial \rho_{\mathrm{n}}},
\end{equation}
\begin{equation}
G_{\tau}^{\mathrm{pp}1} = \frac{b}{\rho_{\mathrm{p}}} \frac{\partial B(\rho_{\mathrm{p}},\rho_{\mathrm{n}})}{\partial \rho_\tau}
+ (1-2\delta_{\tau,\mathrm{p}}) \frac{\partial C(\rho_{\mathrm{p}},\rho_{\mathrm{n}})}{\partial \rho_\tau}
- \delta_{\tau,\mathrm{n}} \frac{\partial D(\rho_{\mathrm{n}},\rho_{\mathrm{p}})}{\partial \rho_{\mathrm{p}}},
\end{equation}
\begin{equation}
G_{\tau}^{\mathrm{np}} = -2 \frac{\partial C(\rho_\tau,\rho_{-\tau})}{\partial \rho_{-\tau}},
\end{equation}
\begin{equation}
G_{\tau}^{\mathrm{nn}2} = c \frac{\partial B(\rho_{\mathrm{n}},\rho_{\mathrm{p}})}{\partial \rho_\tau}
- 2\delta_{\tau,\mathrm{n}} C(\rho_{\mathrm{n}},\rho_{\mathrm{p}})
- \delta_{\tau,\mathrm{p}} D(\rho_{\mathrm{n}},\rho_{\mathrm{p}}),
\end{equation}
\begin{equation}
G_{\tau}^{\mathrm{pp}2} = c \frac{\partial B(\rho_{\mathrm{p}},\rho_{\mathrm{n}})}{\partial \rho_\tau}
- 2\delta_{\tau,\mathrm{p}} C(\rho_{\mathrm{p}},\rho_{\mathrm{n}})
- \delta_{\tau,\mathrm{n}} D(\rho_{\mathrm{n}},\rho_{\mathrm{p}}).
\end{equation}

\section{THE CHARACTERISTIC QUANTITIES OF THE SNM EOS AND THE SYMMETRY ENERGY, THE FOURTH-ORDER SYMMETRY ENERGY, AND THE NUCLEON EFFECTIVE MASSES}
\label{sec:Apd_1}
In this Appendix, we provide the expressions for some quantities appearing in the main text within the MDI3Y model.

Based on Eq.~(\ref{eq:EOS3Y}), the EOS of SNM $E_{0}(\rho)$ can be expressed as
\begin{equation}
\label{eq:E0}
\begin{aligned}
E_{0}(\rho) =& \frac{3 a^{2} \hbar^{2}}{10 m} \rho^{2/3} + (A_{l} + A_{u}) \frac{\rho}{4\rho_{0}} + \frac{t_{3}^{[1]}}{16} \rho^{4/3}
+\frac{t_{3}^{[3]}}{16} \rho^{6/3} +\frac{t_{3}^{[5]}}{16} \rho^{8/3}  \\
&+\frac{1}{12 \pi^{4} \hbar^{6} \rho_{0} \rho}  \sum_{i=1}^{3} \left( C_{l,i}+C_{u,i} \right) \Lambda_{i}^{2}
\left[ p_{F}^{2} (6p_{F}^{2}- \Lambda_{i}^{2}) -8\Lambda_{i} p_{F}^{3} \arctan\frac{2p_{F}}{\Lambda_{i}} + \frac{1}{4} (\Lambda_{i}^{4} + 12 p_{F}^{2} \Lambda_{i}^{2}) \ln \frac{4p_{F}^{2} + \Lambda_{i}^{2}}{\Lambda_{i}^{2}} \right].
\end{aligned}
\end{equation}
%%%%%%%%%%%%%%%%%%%%%%%%%%%%%%
The symmetry energy $E_{\mathrm{sym}}(\rho)$ can be expressed as
\begin{equation}
\label{eq:Esym}
\begin{aligned}
E_{\mathrm{sym}}(\rho) =& \frac{a^2 \hbar^2}{6 m}\rho^{2/3}
+ (A_{l} - A_{u}) \frac{\rho}{4 \rho_0}
- \frac{1}{48} t_{3}^{[1]} \left(2 x_{3}^{[1]} +1 \right) \rho^{4/3}
- \frac{1}{48} t_{3}^{[3]} \left(2 x_{3}^{[3]} +1 \right) \rho^{6/3}
- \frac{1}{48} t_{3}^{[5]} \left(2 x_{3}^{[5]} +1 \right) \rho^{8/3}
\\
&- \frac{1}{6 \pi^2 \hbar^3 \rho_0} \sum_{i=1}^{3} C_{u,i} \, \Lambda_{i}^{2} \, p_{F} \ln \frac{4 p_{F}^{2} + \Lambda_{i}^{2}}{\Lambda_{i}^{2}} + \frac{1}{24 \pi^2 \hbar^3 \rho_0} \sum_{i=1}^{3} \left( C_{l,i}+C_{u,i} \right) \Lambda_{i}^{2} \, p_{F} \left( 4 - \frac{\Lambda_{i}^{2}}{p_{F}^{2}} \ln \frac{4 p_{F}^{2} + \Lambda_{i}^{2}}{\Lambda_{i}^{2}} \right).
\end{aligned}
\end{equation}
%%%%%%%%%%%%%%%%%%%%%%%%%%%%%%

The Taylor expansion coefficients defined in Eqs.~(\ref{eq:L_def})-(\ref{eq:H_def}) are commonly used to characterize the density behaviors of $E_{0}(\rho)$ and $E_{\mathrm{sym}}(\rho)$, which can be expressed as
\begin{equation}
\label{eq:L0}
\begin{aligned}
L_{0}(\rho) =& \frac{3 a^{2} \hbar^{2}}{5 m} \rho^{2/3}
+ (A_{l} + A_{u}) \frac{3 \rho}{4 \rho_{0}}
+ \frac{t_{3}^{[1]}}{4} \rho^{4/3}
+ \frac{3t_{3}^{[1]}}{8} \rho^{6/3}
+ \frac{t_{3}^{[5]}}{2} \rho^{8/3}
\\
&+ \frac{a^{6} \rho}{12 \pi^{4} \rho_{0}} \sum_{i=1}^{3} \left( C_{l,i}+C_{u,i} \right)  \left[  6 \frac{\Lambda_{i}^{2}}{p_{F}^{2}} + 3 \frac{\Lambda_{i}^{4}}{p_{F}^{4}} - \frac{3 \Lambda_{i}^{4} (4 p_{F}^{2} + \Lambda_{i}^{2} )}{4 p_{F}^{6}} \ln \frac{4p_{F}^{2} + \Lambda_{i}^{2}}{\Lambda_{i}^{2}} \right],
\end{aligned}
\end{equation}
%%%%%%%%%%%%%%%%%%%%%%%%%%%%%%
\begin{equation}
\label{eq:K0}
\begin{aligned}
K_{0}(\rho) = & -\frac{3 a^{2} \hbar^{2}}{5 m} \rho^{2/3}
+ \frac{t_{3}^{[1]}}{4} \rho^{4/3}
+ \frac{9t_{3}^{[3]}}{8} \rho^{6/3}
+ \frac{5t_{3}^{[5]}}{2} \rho^{8/3}
\\
& +\frac{a^{9} \hbar^{3} \rho^2}{4 \pi^{4} \rho_{0}} \sum_{i=1}^{3} \left( C_{l,i}+C_{u,i} \right) \left[ -\frac{4 p_{F}^{2} \Lambda_{i}^{2} + 6 \Lambda_{i}^{4} }{p_{F}^{7}} + \frac{\Lambda_{i}^{4}(8p_{F}^{2} + 3\Lambda_{i}^{2})}{2p_{F}^{9}} \ln \frac{4p_{F}^{2} + \Lambda_{i}^{2}}{\Lambda_{i}^{2}}
\right],
\end{aligned}
\end{equation}
%%%%%%%%%%%%%%%%%%%%%%%%%%%%%%
\begin{equation}
\label{eq:J0}
\begin{aligned}
J_{0}(\rho) =& \frac{12 a^2 \hbar^2}{5 m} \rho^{2/3}
- \frac{t_{3}^{[1]}}{2} \rho^{4/3}
+ 5t_{3}^{[5]} \rho^{8/3}
\\
&+ \frac{a^{12} \hbar^{6} \rho^{3}}{4 \pi^{4} \rho_0} \sum_{i=1}^{3} \left( C_{l,i}+C_{u,i} \right) \left[  \frac{2(40 p_{F}^{4} \Lambda_{i}^{2} + 110 p_{F}^{2} \Lambda_{i}^{4} +27 \Lambda_{i}^{6})}{p_{F}^{10}(4 p_{F}^{2} + \Lambda_{i}^{2})} - \frac{56 p_{F}^{2} \Lambda_{i}^{4} + 27 \Lambda_{i}^{6}}{2 p_{F}^{12}} \ln \frac{4p_{F}^{2} + \Lambda_{i}^{2}}{\Lambda_{i}^{2}}
\right],
\end{aligned}
\end{equation}
%%%%%%%%%%%%%%%%%%%%%%%%%%%%%%
\begin{equation}
\label{eq:I0}
\begin{aligned}
I_{0}(\rho) =& -\frac{84 a^2 \hbar ^2}{5 m} \rho^{2/3}
+ \frac{5 t_{3}^{[1]}}{2} \rho^{4/3}
- 5 t_{3}^{[5]} \rho^{8/3} \\
& +\sum_{i=1}^{3} \left( C_{l,i}+C_{u,i} \right) \frac{3\Lambda_{i}^{2}}{4 \pi^2 \hbar^3 \rho_0 p_{F}^{3}} \Bigg[
-4 p_{F}^{2} \frac{32 p_{F}^{6} + 1272 p_{F}^{4} \Lambda_{i}^{2} + 626 p_{F}^{2} \Lambda_{i}^{4} + 81 \Lambda_{i}^{6}}
{\left( 4 p_{F}^{2} + \Lambda_{i}^{2}  \right)^{2} }
+ \left( 140 p_{F}^{2} \Lambda_{i}^{2} + 81 \Lambda_{i}^{4}   \right)
\ln \frac{4p_{F}^{2} + \Lambda_{i}^{2}}{\Lambda_{i}^{2}}
\Bigg],
\end{aligned}
\end{equation}
%%%%%%%%%%%%%%%%%%%%%%%%%%%%%%
\begin{equation}
\label{eq:H0}
\begin{aligned}
H_{0}(\rho) =& \frac{64168 a^2 \hbar ^2}{m} \rho^{2/3}
- 20 t_{3}^{[1]} \rho^{4/3}
+ 20 t_{3}^{[5]} \rho^{8/3} \\
& +\sum_{i=1}^{3} \left( C_{l,i}+C_{u,i} \right) \frac{3\Lambda_{i}^{2}}{4 \pi^2 \hbar^3 \rho_0 p_{F}^{3}} \Bigg[ 4 p_{F}^{2} \frac{14080 p_{F}^{8} + 72864 p_{F}^{6} \Lambda_{i}^{2} + 53840 p_{F}^{4} \Lambda_{i}^{4} + 13970 p_{F}^{2} \Lambda_{i}^{6}
+ 1215 \Lambda_{i}^{8}}
{\left( 4 p_{F}^{2} + \Lambda_{i}^{2}  \right)^{3} } \\
& ~~ -5 \left( 364 p_{F}^{2} \Lambda_{i}^{2} + 243 \Lambda_{i}^{4}\right)
\ln \frac{4p_{F}^{2} + \Lambda_{i}^{2}}{\Lambda_{i}^{2}}
\Bigg],
\end{aligned}
\end{equation}
%%%%%%%%%%%%%%%%%%%%%%%%%%%%%%
and
\begin{equation}
\label{eq:L}
\begin{aligned}
L(\rho)=& \frac{a^2 \hbar^2}{3m} \rho^{2/3}
+ (A_{l} - A_{u}) \frac{3 \rho}{4 \rho_0}
- \frac{1}{12} t_{3}^{[1]} \left(2 x_{3}^{[1]} +1 \right) \rho^{4/3}
- \frac{1}{8} t_{3}^{[3]} \left(2 x_{3}^{[3]} +1 \right) \rho^{6/3}\\
&- \frac{1}{6} t_{3}^{[5]} \left(2 x_{3}^{[5]} +1 \right) \rho^{8/3}
- \frac{\rho}{4\rho_0} \sum_{i=1}^{3} C_{u,i} \left[
\frac{8 \Lambda_{i}^{2}}{4p_{F}^{2} + \Lambda_{i}^{2} } + \frac{\Lambda_{i}^{2}}{p_{F}^{2}} \ln \frac{4 p_{F}^{2} + \Lambda_{i}^{2}}{\Lambda_{i}^{2}}
\right] \\
& + \frac{\rho}{16 \rho_0} \sum_{i=1}^{3} \left( C_{l,i}+C_{u,i} \right) \left[
\frac{4\Lambda_{i}^{2} (4 p_{F}^{2} - \Lambda_{i}^{2})}{p_{F}^{2} (4 p_{F}^{2} + \Lambda_{i}^{2})} + \frac{\Lambda_{i}^{4}}{p_{F}^{4}} \ln \frac{4 p_{F}^{2} + \Lambda_{i}^{2}}{\Lambda_{i}^{2}} \right],
\end{aligned}
\end{equation}
%%%%%%%%%%%%%%%%%%%%%%%%%%%%%%
\begin{equation}
\label{eq:Ksym}
\begin{aligned}
K_{\mathrm{sym}}(\rho)=& -\frac{a^2 \hbar^2}{3 m}\rho^{2/3}
- \frac{1}{12} t_{3}^{[1]} \left(2 x_{3}^{[1]} +1 \right) \rho^{4/3}
- \frac{3}{8} t_{3}^{[3]} \left(2 x_{3}^{[3]} +1 \right) \rho^{6/3}
- \frac{5}{6} t_{3}^{[5]} \left(2 x_{3}^{[5]} +1 \right) \rho^{8/3}
\\
&-\sum_{i=1}^{3} C_{u,i} \frac{p_F \, \Lambda_{i}^{2}}{2 p_{F_0}^{3}} \left[  \frac{4p_{F}^{2}\left(-4p_{F}^{2} + \Lambda_{i}^{2} \right)}{\left(4p_{F}^{2} + \Lambda_{i}^{2} \right)^2} - \ln \frac{4 p_{F}^{2} + \Lambda_{i}^{2}}{\Lambda_{i}^{2}} \right] \\
&+ \sum_{i=1}^{3} \left( C_{l,i}+C_{u,i} \right) \frac{p_F \, \Lambda_{i}^{2}}{4 p_{F_0}^{3}} \left[ \frac{4\left( -8p_{F}^{4} +6p_{F}^{2}\Lambda_{i}^{2} +\Lambda_{i}^{4} \right)}{\left(4p_{F}^{2} + \Lambda_{i}^{2} \right)^2} -\frac{\Lambda_{i}^{2}}{p_{F}^2} \ln \frac{4 p_{F}^{2} + \Lambda_{i}^{2}}{\Lambda_{i}^{2}} \right],
\end{aligned}
\end{equation}
%%%%%%%%%%%%%%%%%%%%%%%%%%%%%%
\begin{equation}
\label{eq:Jsym}
\begin{aligned}
J_{\mathrm{sym}}(\rho)=& \frac{4 a^2 \hbar^2}{3m}\rho^{2/3}
+ \frac{1}{6} t_{3}^{[1]} \left(2 x_{3}^{[1]} +1 \right) \rho^{4/3}
- \frac{5}{3} t_{3}^{[5]} \left(2 x_{3}^{[5]} +1 \right) \rho^{8/3}
\\
&-\sum_{i=1}^{3} C_{u,i} \frac{p_F \, \Lambda_{i}^{2}}{2 p_{F_0}^{3}} \left[ \frac{4 p_{F}^2}{\left(4p_{F}^{2} + \Lambda_{i}^{2} \right)^3} \left(
48 p_{F}^{4} -40 p_{F}^{2} \Lambda_{i}^{2} -5 \Lambda_{i}^{4} \right) + 5 \ln \frac{4 p_{F}^{2} + \Lambda_{i}^{2}}{\Lambda_{i}^{2}} \right] \\
&+ \sum_{i=1}^{3} \left( C_{l,i}+C_{u,i} \right) \frac{p_F \, \Lambda_{i}^{2}}{4 p_{F_0}^{3}} \left[ \frac{4}{\left(4p_{F}^{2} + \Lambda_{i}^{2} \right)^3} \left(
160 p_{F}^{6} -192 p_{F}^{4} \Lambda_{i}^{2} -70 p_{F}^{2} \Lambda_{i}^{4} -7 \Lambda_{i}^{6} \right) +7 \frac{\Lambda_{i}^{2}}{p_{F}^{2}} \ln \frac{4 p_{F}^{2} + \Lambda_{i}^{2}}{\Lambda_{i}^{2}} \right],
\end{aligned}
\end{equation}
%%%%%%%%%%%%%%%%%%%%%%%%%%%%%%
\begin{equation}
\label{eq:Isym}
\begin{aligned}
I_{\mathrm{sym}}(\rho)=& -\frac{28 a^2 \hbar^2}{3m} \rho^{2/3}
- \frac{5}{6} t_{3}^{[1]} \left(2 x_{3}^{[1]} +1 \right) \rho^{4/3}
+ \frac{5}{3} t_{3}^{[5]} \left(2 x_{3}^{[5]} +1 \right) \rho^{8/3}
\\
& -\sum_{i=1}^{3} C_{u,i} \frac{4 p_F \, \Lambda_{i}^{2}}{p_{F_0}^{3}} \left[
\frac{4 p_F^2}{\left(4p_{F}^{2} + \Lambda_{i}^{2} \right)^4} \left( -112 p_{F}^{6} +248 p_{F}^{4} \Lambda_{i}^{2} + 65 p_{F}^{2} \Lambda_{i}^{4} + 5 \Lambda_{i}^{6} \right) -5 \ln \frac{4 p_{F}^{2} + \Lambda_{i}^{2}}{\Lambda_{i}^{2}} \right] \\
& +\sum_{i=1}^{3} \left( C_{l,i}+C_{u,i} \right) \frac{p_F \, \Lambda_{i}^{2}}{2 p_{F_0}^{3}}
\Bigg[  \frac{4}{\left(4p_{F}^{2} + \Lambda_{i}^{2} \right)^3} \left(
-2560 p_{F}^{8} + 4128 p_{F}^{6} \Lambda_{i}^{2} + 2400 p_{F}^{4} \Lambda_{i}^{4} + 490 p_{F}^{2} \Lambda_{i}^{6} + 35 \Lambda_{i}^{8} \right) \\
& ~~ -35 \frac{\Lambda_{i}^{2}}{p_{F}^{2}} \ln \frac{4 p_{F}^{2} + \Lambda_{i}^{2}}{\Lambda_{i}^{2}} \Bigg],
\end{aligned}
\end{equation}
%%%%%%%%%%%%%%%%%%%%%%%%%%%%%%
\begin{equation}
\label{eq:Hsym}
\begin{aligned}
H_{\mathrm{sym}}(\rho)=& \frac{280 a^2 \hbar^2}{3m} \rho^{2/3}
+ \frac{20}{3} t_{3}^{[1]} \left(2 x_{3}^{[1]} +1 \right) \rho^{4/3}
- \frac{20}{3} t_{3}^{[5]} \left(2 x_{3}^{[5]} +1 \right) \rho^{8/3}
\\
& -\sum_{i=1}^{3} C_{u,i} \frac{4 p_F \, \Lambda_{i}^{2}}{p_{F_0}^{3}} \Bigg[ \frac{4 p_F^2}{\left(4p_{F}^{2} + \Lambda_{i}^{2} \right)^5} \left( 2368 p_{F}^{8} -15120 p_{F}^{6} \Lambda_{i}^{2} - 6100 p_{F}^{4} \Lambda_{i}^{4} - 955 p_{F}^{2} \Lambda_{i}^{6} - 55 \Lambda_{i}^{8} \right) + 55 \ln \frac{4 p_{F}^{2} + \Lambda_{i}^{2}}{\Lambda_{i}^{2}} \Bigg] \\
& +\sum_{i=1}^{3} \left( C_{l,i}+C_{u,i} \right) \frac{p_F \, \Lambda_{i}^{2}}{2 p_{F_0}^{3}}
\Bigg[ \frac{4}{\left(4p_{F}^{2} + \Lambda_{i}^{2} \right)^5} ( 112640 p_{F}^{10} -224896 p_{F}^{8} \Lambda_{i}^{2} -182560 p_{F}^{6} \Lambda_{i}^{4} -56840 p_{F}^{4} \Lambda_{i}^{6} \\ & ~~ -8190 p_{F}^{2} \Lambda_{i}^{8} -455 \Lambda_{i}^{10} )
+455 \frac{\Lambda_{i}^{2}}{p_{F}^{2}} \ln \frac{4 p_{F}^{2} + \Lambda_{i}^{2}}{\Lambda_{i}^{2}}
\Bigg].
\end{aligned}
\end{equation}
%%%%%%%%%%%%%%%%%%%%%%%%%%%%%%

The fourth-order symmetry energy can be expressed as
\begin{equation}
\label{eq:Esym4}
\begin{aligned}
E_{\mathrm{sym},4}(\rho) \equiv \left. \frac{1}{4!} \frac{\partial^{4} E(\rho,\delta)}{\partial \delta^{4}} \right|_{\delta=0}
=& \frac{a^{2} \hbar ^2}{162m} \rho^{2/3} + \sum_{i=1}^{3} \frac{p_{F}}{108 p_{F_0}^{3}} \Bigg[ C_{u,i}  \left( 2p_{F}^{2} + 7\Lambda_{i}^{2} \right) +C_{l,i} \Lambda_{i}^{2}  \frac{40p_{F}^{4} + 42p_{F}^{2} \Lambda_{i}^{2} +7\Lambda_{i}^4 }{\left( 4p_{F}^{2} + \Lambda_{i}^{2} \right)^{2}} \\
&-\frac{\Lambda_{i}^{2}}{4p_{F}^2} \left(16 C_{u,i} p_{F}^{2} +7C_{l,i} \Lambda_{i}^{2} + 7C_{u,i} \Lambda_{i}^{2} \right)  \ln \frac{4 p_{F}^{2} + \Lambda_{i}^{2}}{\Lambda_{i}^{2}} \Bigg].
\end{aligned}
\end{equation}

In the MDI3Y model, the nucleon effective mass can be generally expressed as
\begin{equation}
\label{eq:mtau_exp}
\begin{aligned}
\left[\frac{m_{\tau}(\rho,\delta,p)}{m}\right]^{-1} \equiv & 1 + \frac{m}{p} \frac{\mathrm{d}U_{\tau}(\rho,\delta,p)}{\mathrm{d}p} \\
=& 1 + \frac{m}{4 \pi^2 \hbar^3 \rho_0} \sum_{i=1}^{3} C_{l,i} \frac{\Lambda_{i}^{2}}{ p^3} \left[ 4 p \, p_{F,\tau} - (p^2 + p_{F,\tau}^{2} + \Lambda_{i}^{2}) \ln \frac{(p+p_{F,\tau})^2 +\Lambda_{i}^{2}}{(p-p_{F,\tau})^2 +\Lambda_{i}^{2}} \right] \\
&+ \frac{m}{4 \pi^2 \hbar^3 \rho_0} \sum_{i=1}^{3} C_{u,i} \frac{\Lambda_{i}^{2}}{ p^3} \left[ 4 p \, p_{F,-\tau} - (p^2 + p_{F,-\tau}^{2} + \Lambda_{i}^{2}) \ln \frac{(p+p_{F,-\tau})^2 +\Lambda_{i}^{2}}{(p-p_{F,-\tau})^2 +\Lambda_{i}^{2}} \right].
\end{aligned}
\end{equation}
The isoscalar nucleon effective mass $m_{s}^{\ast}$ is the nucleon effective mass in SNM, i.e.,
\begin{equation}
\label{eq:ms_exp}
\begin{aligned}
\left( \frac{m_{s}^{\ast}}{m} \right)^{-1} \equiv &  1 + \frac{m}{p} \frac{ \mathrm{d}U_{0}(\rho,p)}{\mathrm{d}p} \bigg|_{p=p_{F,\tau}=p_F} \\
=& 1 + \frac{m}{4 \pi^{2} \hbar^{3} \rho_{0}} \sum_{i=1}^{3} \left( C_{l,i}+C_{u,i} \right)  \frac{\Lambda_{i}^{2}}{p_{F}^3}   \left[4 p_{F}^2 - (2 p_{F}^2 + \Lambda_{i}^{2} ) \ln \frac{4 p_{F}^2 +\Lambda_{i}^{2}}{\Lambda_{i}^{2}} \right],
\end{aligned}
\end{equation}
while the isovector nucleon effective mass $m_{v}^{\ast}$ equals to the effective mass of proton (neutron) in pure neutron (proton) matter, i.e.,
\begin{equation}
\label{eq:mv}
\begin{aligned}
\left(\frac{m^{\ast}_{v}}{m}\right)^{-1} \equiv 1 + \frac{m}{p} \frac{\mathrm{d} U_{\tau}(\rho,\delta=-\tau,p)}{\mathrm{d}p} \bigg|_{p=p_{F,\tau}=0}
= 1- \frac{8 m}{3\pi^2 \hbar^3 \rho_0} \sum_{i=1}^{3} C_{u,i} \frac{p_{F}^{3} \Lambda_{i}^{2}}{(2^{2/3} p_{F}^{2} +\Lambda_{i}^{2} )^{2}}.
\end{aligned}
\end{equation}
The linear isospin splitting coefficient $\Delta m^{\ast}_{1}(\rho)$ can be expressed as
\begin{equation}
\label{eq:Dm1}
\begin{aligned}
\Delta m^{\ast}_{1}(\rho) \equiv \left. \frac{\partial \msplast(\rho,\delta)}{\partial \delta} \right|_{\delta=0} = \frac{8}{3} p_{F}^{5} \frac{ \mathcal{A}(\rho) }{ \left[ \mathcal{B}(\rho) \right]^2 },
\end{aligned}
\end{equation}
where
\begin{equation}
\label{eq:Arho}
\begin{aligned}
\mathcal{A}(\rho) = \sum_{i=1}^{3} \Lambda_{i}^{2} \left\{ \left( 2\Tilde{C}_{l,i} +3\Tilde{C}_{u,i} \right) + \Tilde{C}_{l,i} \frac{\Lambda_{i}^{2}}{4p_{F}^{2}+\Lambda_{i}^{2}}
-\left[ \frac{1}{2} \left( \Tilde{C}_{l,i} +3\Tilde{C}_{u,i} \right) + \frac{3}{4} \left( \Tilde{C}_{l,i} +\Tilde{C}_{u,i} \right) \frac{\Lambda_{i}^{2}}{p_{F}^{2}} \right] \ln \frac{4 p_{F}^{2} + \Lambda_{i}^{2}}{\Lambda_{i}^{2}} \right\},
\end{aligned}
\end{equation}
%%%%%%%%%%%%%%%%%%%%%%%%%%%%%%
and
\begin{equation}
\label{eq:Brho}
\begin{aligned}
\mathcal{B}(\rho) = p_{F}^{3} + \sum_{i=1}^{3} \Lambda_{i}^{2} \left[ 4(\Tilde{C}_{l,i}+\Tilde{C}_{u,i}) p_{F}^{2} -(\Tilde{C}_{l,i}+\Tilde{C}_{u,i}) \left( 2p_{F}^{2} + \Lambda_{i}^{2} \right) \ln \frac{4 p_{F}^{2} + \Lambda_{i}^{2}}{\Lambda_{i}^{2}} \right],
\end{aligned}
\end{equation}
with $\Tilde{C}_{l,i}=\frac{3m}{8 p_{F_0}^{3}} C_{l,i}$ and $\Tilde{C}_{u,i}=\frac{3m}{8 p_{F_0}^{3}} C_{u,i}$.

\end{widetext}

\bibliography{MDI3Y}

\end{document}